\documentclass[twocolumn]{aastex701}
\usepackage{threeparttable}
\usepackage{multirow}
\usepackage{adjustbox}
\usepackage{graphicx}
\usepackage{booktabs} 
\usepackage{enumitem}
\usepackage{hhline}
\usepackage{caption}
\usepackage{hyperref}
\usepackage{textcomp}
\newcommand{\chandra}{{Chandra}}
\newcommand\iona[2]{#1$\;${\scshape{#2}}}
\begin{document}
\title{Deep X-ray View of a Distant Protocluster Anchored by a Quasar at $z=6.63$: No Evidence for Enhanced X-ray AGN Activity}
\correspondingauthor{Shouyi Wang} \email {sywang0302@gmail.com}
\author[0009-0003-1260-4143]{Shouyi Wang} \email{sywang0302@gmail.com}
\affiliation{Department of Astronomy, School of Physics, Peking University, Beijing 100871, People's Republic of China}
\affiliation{Kavli Institute for Astronomy and Astrophysics, Peking University, Beijing 100871, People's Republic of China}
\affiliation{School of Astronomy and Space Science, Nanjing University, Nanjing, Jiangsu 210093, People’s Republic of China}
\affiliation{Key Laboratory of Modern Astronomy and Astrophysics (Nanjing University), Ministry of Education, Nanjing 210093, People’s Republic of China}
\author[0000-0002-7633-431X]{Feige Wang} \email {fgwang@umich.edu}
\affiliation{Department of Astronomy, University of Michigan, 1085 South University Avenue, Ann Arbor, MI 48109, USA}
\author[0000-0001-5287-4242]{Jinyi Yang} \email{}
\affiliation{Department of Astronomy, University of Michigan, 1085 South University Avenue, Ann Arbor, MI 48109, USA}
\author[0000-0002-9036-0063]{Bin Luo} \email {bluo@nju.edu.cn}
\affiliation{School of Astronomy and Space Science, Nanjing University, Nanjing, Jiangsu 210093, People’s Republic of China}
\affiliation{Key Laboratory of Modern Astronomy and Astrophysics (Nanjing University), Ministry of Education, Nanjing 210093, People’s Republic of China}
\author[0000-0002-5768-738X]{Xiangyu Jin} \email{}
\affiliation{Department of Astronomy, University of Michigan, 1085 South University Avenue, Ann Arbor, MI 48109, USA}

\author[0000-0002-4436-6923]{Fan Zou}\email {fanzou01@gmail.com}
\affiliation{Department of Astronomy, University of Michigan, 1085 South University Avenue, Ann Arbor, MI 48109, USA}
\author[0000-0002-6184-9097]{Jaclyn B. Champagne}
\email{jchampagne@stsci.edu}
\affiliation{Space Telescope Science Institute, 3700 San Martin Drive, Baltimore, MD 21218, USA}
\author[0000-0003-3310-0131]{Xiaohui Fan} \email{}
\affiliation{Steward Observatory, University of Arizona, 933 N Cherry Avenue, Tucson, AZ 85721, USA}
\author{Elaina Malczewski} \email{}
\affiliation{Department of Astronomy, University of Michigan, 1085 South University Avenue, Ann Arbor, MI 48109, USA}
\author[0000-0003-0680-9305]{Fabio Vito} \email{fabio.vito@inaf.it}
\affiliation{INAF -- Osservatorio di Astrofisica e Scienza dello Spazio di Bologna, Via Gobetti 93/3, I-40129 Bologna, Italy}
\author[0000-0002-7350-6913]{Xue-Bing Wu} \email{}
\email{wuxb@pku.edu.cn}
\affiliation{Department of Astronomy, School of Physics, Peking University, Beijing 100871, People's Republic of China}
\affiliation{Kavli Institute for Astronomy and Astrophysics, Peking University, Beijing 100871, People's Republic of China}

\author[0000-0002-2931-7824]{Eduardo Ba\~nados} \email{}
\affiliation{Max Planck Institut f\"ur Astronomie, K\"onigstuhl 17, D-69117 Heidelberg, Germany}

\author[0000-0002-1620-0897]{Fuyan Bian} \email{}
\affiliation{European Southern Observatory, Alonso de C\'ordova 3107, Casilla 19001, Vitacura, Santiago 19, Chile}

\author[0000-0002-7898-7664]{Thomas Connor}
\email{thomas.connor@cfa.harvard.edu}
\affiliation{Center for Astrophysics $\vert$\ Harvard\ \&\ Smithsonian, 60 Garden St., Cambridge, MA 02138, USA}

\author[0000-0002-6748-2900]{Tiago Costa} \email{}
\affiliation{School of Mathematics, Statistics and Physics, Newcastle University, Newcastle upon Tyne, NE1 7RU, UK}

\author[0000-0003-2895-6218]{Anna-Christina Eilers} \email{}
\affiliation{Department of Physics, Massachusetts Institute of Technology, Cambridge, MA 02139, USA}
\affiliation{MIT Kavli Institute for Astrophysics and Space Research, Massachusetts Institute of Technology, Cambridge, MA 02139, USA}

\author[0000-0002-6822-2254]{Emanuele Paolo Farina} \email{}
\affiliation{International Gemini Observatory, NSF NOIRLab, 670 N A`ohoku Place, HI-96720, Hilo, USA}
\affiliation{INAF -- Osservatorio di Astrofisica e Scienza dello Spazio di Bologna, via Gobetti 93/3, I 40129, Bologna, Italy}
\author[0000-0002-7054-4332]{Joseph F.\ Hennawi} \email{}
\affiliation{Department of Physics, University of California, Santa Barbara, CA 93106-9530, USA}
\affiliation{Leiden Observatory, Leiden University, Niels Bohrweg 2, NL-2333 CA Leiden, Netherlands}

\author[0000-0003-1470-5901]{Hyunsung D. Jun} \email{}
\affiliation{Department of Physics, Northwestern College, 101 7th St SW, Orange City, IA 51041, USA}
\affiliation{School of Physics, Korea Institute for Advanced Study, 85 Hoegiro, Dongdaemun-gu, Seoul 02455, Republic of Korea}

\author[0000-0001-6239-3821]{Jiang-Tao Li} \email{}
\affiliation{Purple Mountain Observatory, Chinese Academy of Sciences, 10 Yuanhua Road, Nanjing 210023, People's Republic of China}

\author[0000-0002-5941-5214]{Chiara Mazzucchelli}\email{}
\affiliation{Instituto de Estudios Astrof\'{\i}sicos, Facultad de Ingenier\'{\i}a y Ciencias, Universidad Diego Portales, Avenida Ejercito Libertador 441, Santiago, Chile}

\author[0009-0008-2205-7725]{Klaudia Protušová} \email{}
\affiliation{Institute for Theoretical Physics, Heidelberg University, Philosophenweg 12, 69120, Heidelberg, Germany}

\author[0000-0003-4793-7880]{Fabian Walter} \email{}
\affiliation{Max Planck Institut f\"ur Astronomie, K\"onigstuhl 17, D-69117 Heidelberg, Germany}

\author[0000-0002-6990-9058]{Zhibo Yu}\email{}
\affiliation{Department of Astronomy and Astrophysics, 525 Davey Lab, The Pennsylvania State University, University Park, PA 16802, USA}
\affiliation{Institute for Gravitation and the Cosmos, The Pennsylvania State University, University Park, PA 16802, USA}
\author[0000-0002-5367-8021]{Minghao Yue} \email{}
\affiliation{Steward Observatory, University of Arizona, 933 N Cherry Ave, Tucson, AZ 85721, USA}

\begin{abstract}
Whether dense environments accelerate SMBH growth during cosmic reionization remains uncertain because X-ray constraints on $z>6$ protoclusters are rare. We analyze $\sim865$~ks of \chandra\ observations of J0910$-$0414 quasar field, a massive protocluster at $z=6.63$. We construct a catalog of 251 X-ray sources and search for AGN among 21 spectroscopically confirmed members, excluding the central quasar. This is the first systematic X-ray AGN census in a $z>6$ protocluster; none of the members is individually detected. Stacking all members ($\sim17$~Ms total exposure) yields no significant signal and limits the average rest-frame 2--10~keV luminosity to $\langle L_{\rm 2-10~keV}\rangle\leq9.54\times10^{42}~\rm erg~s^{-1}$. The null result gives a central 90\% Jeffreys upper bound of $f_{\rm AGN}<0.09$. After accounting for the spatially varying sensitivity, the field Compton-thin X-ray luminosity function predicts $N_{\rm exp,field}=0.0356$ AGN in the surveyed volume. Assuming the overdensity profile of AGNs in this protocluster is the same as that of galaxies determined by \cite{Wang2024ApJ...962L..11W}, the expected AGN count becomes $N_{\rm exp}=0.131$. Zero detections then imply a one-sided 90\% upper limit of $\eta_{\rm rel}<17.5$ on any additional X-ray AGN enhancement beyond that associated with the galaxy overdensity. Although $\eta_{\rm rel}<17.5$ permits more than an order-of-magnitude enhancement, the data show no significant excess of X-ray-detectable AGN in J0910$-$0414 protocluster. The present X-ray data may miss heavily obscured, intrinsically X-ray-weak, or intermittently accreting SMBHs; the non-detection therefore does not rule out AGNs among the member galaxies.
\end{abstract}
\section{Introduction}

The existence of supermassive black holes (SMBHs) with masses exceeding $10^9~M_{\odot}$ at $z\geq6$ \citep[e.g.,][]{Wu2015Natu,Banados2016ApJS,Banados2023ApJS,Yang2019AJ,Wang2021ApJ,Belladitta2025A&A} challenges our understanding of early black hole growth. In the hierarchical structure formation framework, luminous high-redshift quasars are expected to inhabit rare, massive dark matter halos and have therefore been proposed as signposts of early density peaks that may evolve into massive galaxy clusters (i.e., protoclusters) \citep{Angulo2012MNRAS,Costa2014MNRAS,Matteo2017MNRAS,Lupi2024A&A,Wang2026ApJ}.
\par
Over the past two decades, numerous studies have searched for galaxy overdensities around $z>6$ quasars \citep[e.g.,][]{Stiavelli2005ApJ,Willott2005ApJ,Zheng2006ApJ,Kim2009ApJ,Banados2013ApJ,Morselli2014A&A,Champagne2023ApJ}. Several individual fields have been spectroscopically confirmed to host megaparsec-scale overdensities, including J1030+0524 ($z=6.31$) \citep{Mignoli2020A&A}, J0100+2802 ($z=6.33$) \citep{Kashino2023ApJ}, J1526$-$2050 ($z=6.59$) \citep{Meyer2022ApJ}, J0305$-$3150 ($z=6.61$) \citep{Farina2017ApJ,Wang2023ApJ,Champagne2025ApJ}, and J0910$-$0414 ($z=6.63$) \citep{Wang2024ApJ...962L..11W}. However, larger and more uniform surveys reveal substantial field-to-field variation. The A SPectroscopic survey of biased halos In the Reionization Era (ASPIRE) program, a wide-field slitless spectroscopy (WFSS) grism survey with the Near-Infrared Camera (NIRCam) on the James Webb Space Telescope (JWST), studied the environments of 25 quasars at $z>6.5$. Companion [\iona{O}{iii}] emitters were selected within $|\Delta v_{\rm los}|<1000~{\rm km~s^{-1}}$ and the uniformly covered NIRCam Module A footprint, corresponding to an effective radius of $r_{\rm eff}\lesssim3~{\rm cMpc}$ and a volume of $V\sim500~{\rm cMpc^3}$ per field \citep{Wang2026ApJ}. The number of companions ranges from 0 to 20. Of these 25 quasars, 18 reside in environments denser than the blank field, and seven satisfy the protocluster criterion $\delta>5$, where $\delta\equiv n_{\rm comp}/\bar{n}-1$, with $n_{\rm comp}$ denoting the companion-galaxy number density in a quasar field and $\bar{n}$ the corresponding mean field number density. Similarly, \citet{Eilers2024ApJ} found environments ranging from an extreme overdensity of $\delta\approx65$ within a radius of $2~{\rm cMpc}$ to a density consistent with the mean field across four $z\gtrsim6$ quasar fields. Comparisons with cosmological simulations indicate that this variation reflects intrinsically diverse quasar environments rather than observational bias; luminous $z\sim7$ quasars nevertheless trace galaxy overdensities on average \citep{Wang2026ApJ}.
\par
Whether such overdense environments enhance SMBH growth remains uncertain. High galaxy densities may increase the frequency of interactions and mergers. These events can drive cold gas toward galactic nuclei and trigger star formation and SMBH accretion \citep{Springel2005Natur,Hopkins2008ApJS,Liu2023MNRAS}, while abundant cold-gas reservoirs may further sustain rapid growth \citep{Umehata2019Sci}. However, nuclear accretion is episodic, may occur during heavily obscured phases \citep{Hopkins2006ApJS}, and its observed incidence depends on the luminosity threshold and selection method \citep[e.g.,][]{Hickox2014ApJ,Blecha2018MNRAS}. Rest-frame optical spectroscopy has also been used to search for AGN in $z>6$ protoclusters. In the J0305$-$3150 protocluster at $z=6.61$, \citet{Champagne2025ApJII} found no evidence for broad-line AGN in individual or stacked spectra, while four protocluster members on the outskirts satisfied both the mass--excitation and ``little red dot'' color/compactness criteria and were therefore regarded only as tentative AGN candidates.
\par
Compared with rest-frame UV and optical diagnostics, X-ray observations are more effective in detecting SMBH accretion even in the presence of heavy obscuration \citep{Brandt2015A&ARv}. 

At $z\approx2$--$4.3$, dedicated X-ray studies have reported elevated AGN fractions in several protoclusters \citep[e.g.,][]{Lehmer2009ApJ,Lehmer2013ApJ,Digby2010MNRAS,Tozzi2022A&A,Vito2024A&A,Traina2025A&A,IslaLlave2026arXiv} and have uncovered obscured accretion consistent with rapid, dust-enshrouded SMBH growth \citep{Vito2020A&A,Monson2023ApJ}, although the inferred fractions depend on heterogeneous luminosity thresholds and galaxy selections. \chandra, XMM--Newton, and \textit{Swift} observations have characterized the X-ray emission, spectral properties, and variability of  quasars at $z\gtrsim4.5$ \citep[e.g.,][]{Nanni2017A&A,Vito2019A&A,Pons2020MNRAS,Li2021MNRAS,Li2021ApJ_XrayBright,Zappacosta2023A&A,Hu2025ApJ,Jin2026Apj}. At $z>6$, X-ray studies of quasar environments have so far been limited mainly to individual quasar--companion systems \citep[e.g.,][]{Connor2020ApJ}. However, the X-ray AGN population among member galaxies of $z>6$ protoclusters remains largely unconstrained.
\par
We therefore study the environment of the luminous quasar J0910$-$0414, discovered by \citet{Wang2019ApJ}. {The quasar has a redshift of $z=6.6363\pm0.0003$ measured from the [\iona{C}{ii}] $158~\mu\mathrm{m}$ line \citep{Wang2024ApJ}.} It has a virial black hole mass of $(3.59\pm0.61)\times10^9~M_{\odot}$ \citep{Yang2021ApJ}. \citet{Wang2024ApJ...962L..11W} spectroscopically confirmed that it anchors a massive protocluster with an estimated descendant mass of $M_{z=0}\approx6.9\times10^{15}~M_{\odot}$. The structure has an overall galaxy overdensity of $\delta=3.3^{+1.1}_{-0.9}$ across $\approx35\times74~\rm cMpc^2$ and an extreme central overdensity of $\delta>30$ at $R\lesssim2~\rm cMpc$. It was initially traced by three [\iona{C}{II}] emitters, hereafter referred to as submillimeter galaxies (SMGs), and 12 spectroscopically confirmed Ly$\alpha$ emitters (LAEs) at $6.54<z\leq6.64$. Subsequent JWST/NIRCam WFSS observations identified 32 [\iona{O}{iii}]+H$\beta$ emitters at $5.43<z<6.67$, including 17 within the adopted protocluster redshift interval $6.53<z<6.73$ before the X-ray-specific off-axis and extraction-overlap cuts adopted in this work \citep{Wang2026ApJ}; follow-up JWST/NIRSpec spectroscopy further identified more member galaxies (Malczewski et al. in prep.), yielding a total of 21 galaxies at $6.53<z<6.73$. 

\par
In this work, we present new, deep \chandra\ observations of the J0910$-$0414 field with a total cleaned exposure of $\sim865$~ks and construct a deep X-ray point-source catalog. Our primary environmental test is to identify X-ray AGN and constrain their incidence among the 21 protocluster member galaxies; the central quasar is excluded from the AGN-incidence calculation and is retained only in the field catalog for completeness. We compare these constraints with measurements of lower-redshift protoclusters and with sensitivity-weighted predictions from the field X-ray luminosity function to investigate the influence of overdense environments on early SMBH growth. A detailed analysis of the central quasar's X-ray properties is beyond the scope of this work and will be presented in a separate paper. Throughout this paper, we use J2000 coordinates and adopt a $\Lambda$CDM cosmology with $H_0=70~\mathrm{km~s^{-1}~Mpc~^{-1}}$, $\Omega_\Lambda=0.7$, and $\Omega_M=0.3$.

{The remainder of this paper is organized as follows. Section~\ref{sec:2} summarizes the \chandra\ observations, X-ray source catalog, and survey sensitivity. Section~\ref{sec:4} analyzes the X-ray properties and AGN incidence of the protocluster members and interprets the non-detections. Section~\ref{sec:5} summarizes our conclusions. Appendix~\ref{app:catalog} presents the detailed X-ray survey and catalog construction, and Appendix~\ref{app:specfit} presents the spectral analyses of four hard X-ray sources.}

\section{Chandra observations, source catalog, and survey sensitivity}
\label{sec:2}

{This section summarizes the \chandra\ observations, data reduction, X-ray source catalog, and survey sensitivity. The detailed procedures and supporting material for this section are presented in Appendix~\ref{app:catalog}.}

\subsection{Data reduction}
\label{sec:data_reduction}

The J0910$-$0414 protocluster was observed with \chandra\ 45 times between June 2022 and December 2023, totaling an exposure time of 870~ks. All observations were obtained using the Advanced CCD Imaging Spectrometer imaging array \citep[ACIS-I;][]{Garmire2003SPIE.4851...28G} in VFAINT telemetry mode to minimize particle background and enhance the sensitivity for weak source detection. The ACIS-S2 CCD was additionally active during 24 of the observations, but only events recorded on the four ACIS-I CCDs (CCD IDs 0--3) were used in the analysis. The total survey area covers approximately $515~\mathrm{arcmin^2}$, with individual exposures ranging from 9 to 33~ks. The exposure-weighted mean aim point for the merged dataset is $(\mathrm{R.A.},\mathrm{Decl.})=(137.730^\circ,-4.239^\circ)$.

We reprocessed the data with the \textit{chandra\_repro} script in CIAO 4.17 \citep{CIAO2006SPIE.6270E..1VF}, enabling the \texttt{check\_vf\_pha=yes} option for the VF mode data, and retained only events with ASCA grades 0, 2, 3, 4, and 6. We then ran the \textit{deflare} script and used an iterative $3\sigma$ clipping algorithm to filter potential background flares. The observations were not significantly affected by background flares, and the total cleaned exposure is 865.5~ks. We corrected the relative astrometry against bright sources in the DESI Legacy Imaging Survey DR10 catalog \citep{DESI2019AJ....157..168D}.

\begin{figure*}[htbp]
\centering
\includegraphics[width=\textwidth]{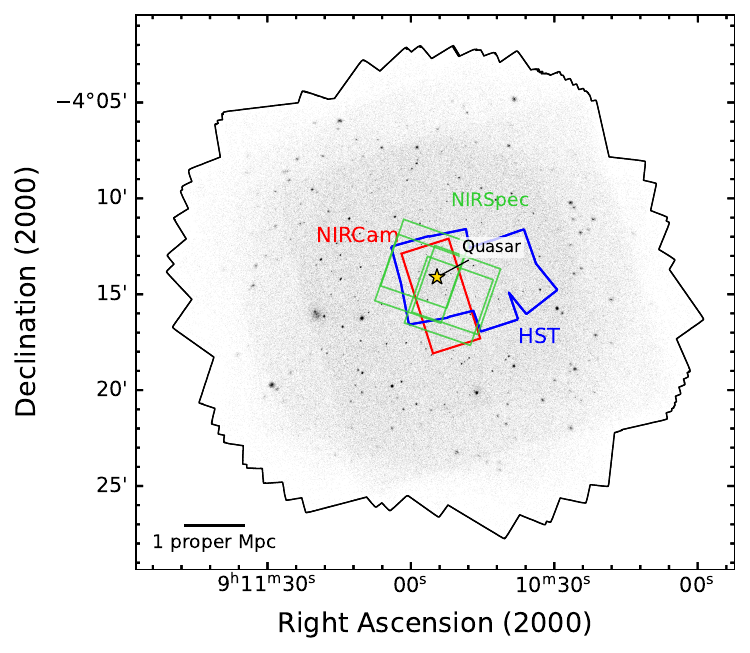}
\caption{Full-band raw image shown in linear grayscale. The black outline delineates the \chandra\ coverage of the J0910$-$0414 field. Blue and red outlines denote the HST/WFC3 and JWST/NIRCam imaging footprints, respectively, while the four green outlines show the JWST/NIRSpec MSA footprints of the two PRISM and two G395M visits. The gold star marks the central quasar. {The HSC mosaics fully encompass the \chandra\ coverage and are omitted for clarity.} \label{fig:ximage}}
\end{figure*}

Using the astrometrically corrected event files, we created merged event files, images, exposure maps, and PSF maps in the full (0.5--7~keV), soft (0.5--2~keV), and hard (2--7~keV) bands. Figure~\ref{fig:ximage} shows the full-band image and the footprints of the HST/WFC3, JWST/NIRCam, and JWST/NIRSpec observations. {The three stitched HSC mosaics in the $i2$, $z$, and NB0926 bands cover approximately $2.06^\circ\times1.88^\circ$ and fully encompass the \chandra\ coverage; their boundary is therefore omitted.} To verify the astrometric accuracy of the merged dataset, we cross-matched the full-band point sources with the \textit{Gaia} DR3 catalog \citep{Gaia2023A&A...674A...1G}. This procedure yielded 18 unique one-to-one matches within $1\arcsec$. The median radial separation is $0.26\arcsec$, with 68\% of the matches lying within $0.44\arcsec$.

\begin{figure*}[htbp]
\centering
\includegraphics[width=\textwidth]{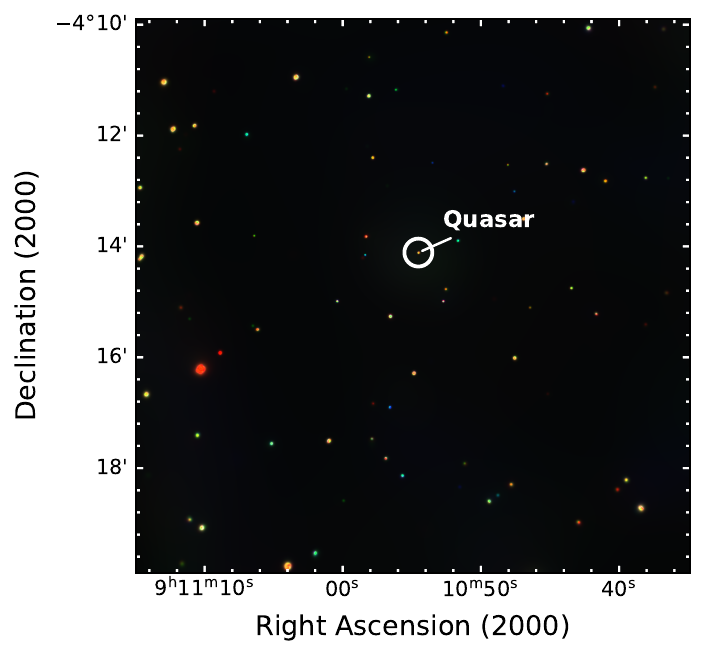}
\caption{Smoothed false-color image of the central $5\arcmin\times5\arcmin$ region. Red, green, and blue correspond to 0.5--2, 2--4, and 4--7~keV, respectively. {The image provides a visual overview of the spatial distribution of the detected sources; no quantitative measurements are derived from this smoothed image.} \label{fig:x3color}}
\end{figure*}

We also generated smoothed images in the 0.5--2~keV, 2--4~keV, and 4--7~keV bands following the method in Section~3.1 of \citet{Luo2017ApJS..228....2L}, using the CIAO tool \textit{csmooth}. These images were then combined into a color composite with the \textit{make\_lupton\_rgb} function in \texttt{Astropy} \citep{Astropy2022ApJ...935..167A}. We show the central $5\arcmin\times5\arcmin$ region in Figure~\ref{fig:x3color}. {The image provides a visual overview of the spatial distribution; it is not used for source detection, photometry, or quantitative spectral analysis.}

\subsection{X-ray source catalog}
\label{sec:3}

To generate the preliminary candidate source list, we applied \textit{wavdetect} to the merged raw images in the full, soft, and hard bands using a false-positive probability threshold of $10^{-6}$. The resulting source lists from the three bands were then cross-matched with a matching radius of $3\arcsec$ to produce a merged source list. For sources detected in multiple bands, the X-ray positions were adopted from the full-band, soft-band, or hard-band detections in order of priority. This procedure yielded a preliminary list of 262 candidate sources. We used \textit{srcflux} to extract the X-ray photometric properties of these sources and retained sources with a binomial no-source probability of $P_{\rm B}<0.01$ in at least one band. Our final catalog consists of 251 sources detected in one or more X-ray bands, of which 247 are detected in the full band.

For sources detected in any of the three X-ray bands, we calculated the uncertainties on their net counts in each band following \citet{Gehrels1986ApJ...303..336G}. If a source was not detected in a given band, we assigned a 90\% confidence-level upper limit to its net counts in that band using the Bayesian method of \citet{Kraft1991ApJ...374..344K}. We converted the count rates or their upper limits to fluxes by adopting an absorbed power-law model with the Galactic column density fixed at $N_{\rm H}=2.7\times10^{20}~{\rm cm}^{-2}$ \citep{HI4PI2016A&A...594A.116H} and the photon index fixed at $\Gamma=1.4$ for all sources.

We searched for ultraviolet, optical, and infrared counterparts to the X-ray sources using the \texttt{NWAY} code \citep{Salvato2018MNRAS.473.4937S} and the \textit{GALEX}, DESI Legacy, HSC, VHS, JWST/NIRCam, and HST/WFC3 catalogs. In the first matching step, 227 X-ray sources (90.4\%) were found to have at least one reliable counterpart. For the remaining 24 sources, a secondary matching process with relaxed criteria identified plausible counterparts for 20 sources. The remaining four X-ray sources have no identified UV, optical, and infrared counterpart. The full 92-column catalog and the merged X-ray images and exposure maps are available on Zenodo at \dataset[doi:10.5281/zenodo.21945827]{https://doi.org/10.5281/zenodo.21945827}.

\subsection{Survey sensitivity}

We constructed sensitivity maps from the source-masked images with the CIAO tool \textit{lim\_sens}, using the spatially varying merged PSF and exposure maps. We set \texttt{snr\_limit}=2.6 and \texttt{bscale}=5; this threshold provides an approximate empirical match to the $P_{\rm B}<0.01$ catalog selection, although the two statistics are not identical. We restricted the calculation to the common $0.1093~\mathrm{deg}^{2}$ region where the full-band exposure exceeds 10\% of its maximum. The deepest flux limits are $1.34\times10^{-16}$, $5.60\times10^{-17}$, and $1.74\times10^{-16}~\mathrm{erg~s^{-1}~cm^{-2}}$ in the full, soft, and hard bands, respectively. At $z=6.63$, the deepest full-band limit corresponds to a rest-frame 2--10~keV luminosity of $1.93\times10^{43}~\mathrm{erg~s^{-1}}$ for $\Gamma=1.4$. Table~\ref{tbl:sensitivity} summarizes the deepest limit and the mean limits within radii of $1\arcmin$ and $8\arcmin$.

\begingroup
\renewcommand{\arraystretch}{1.3}
\begin{deluxetable*}{lcccc}
\tablewidth{0pt}
\tablecaption{2.6$\sigma$ flux and luminosity limits\label{tbl:sensitivity}}
\tablehead{
\colhead{Region} & \colhead{$f_{\mathrm{0.5-7~keV}}$} & \colhead{$f_{\mathrm{0.5-2~keV}}$} & \colhead{$f_{\mathrm{2-7~keV}}$} & \colhead{$L_{\mathrm{2-10~keV}}$} \\[-5pt]
\colhead{} & \multicolumn{3}{c}{$(\mathrm{erg\,s^{-1}\,cm^{-2}})$} & \colhead{$(\mathrm{erg\,s^{-1}})$}
}
\startdata
Deepest & $1.34\times10^{-16}$ & $5.60\times10^{-17}$ & $1.74\times10^{-16}$ & $1.93\times10^{43}$ \\
$r<1\arcmin$ & $1.56\times10^{-16}$ & $5.96\times10^{-17}$ & $2.05\times10^{-16}$ & $2.24\times10^{43}$ \\
$r<8\arcmin$ & $3.55\times10^{-16}$ & $1.16\times10^{-16}$ & $3.98\times10^{-16}$ & $5.09\times10^{43}$ \\
\enddata
\tablecomments{Sensitivity limits derived with \textit{lim\_sens} using \texttt{snr\_limit}=2.6. The luminosity limits are calculated at $z=6.63$ from the full-band values assuming $\Gamma=1.4$. ``Deepest'' is the minimum within the common exposure footprint; the other rows give the mean within the stated radius.}
\end{deluxetable*}
\endgroup

\begin{figure*}[htbp]
\centering
\includegraphics[width=0.49\textwidth]{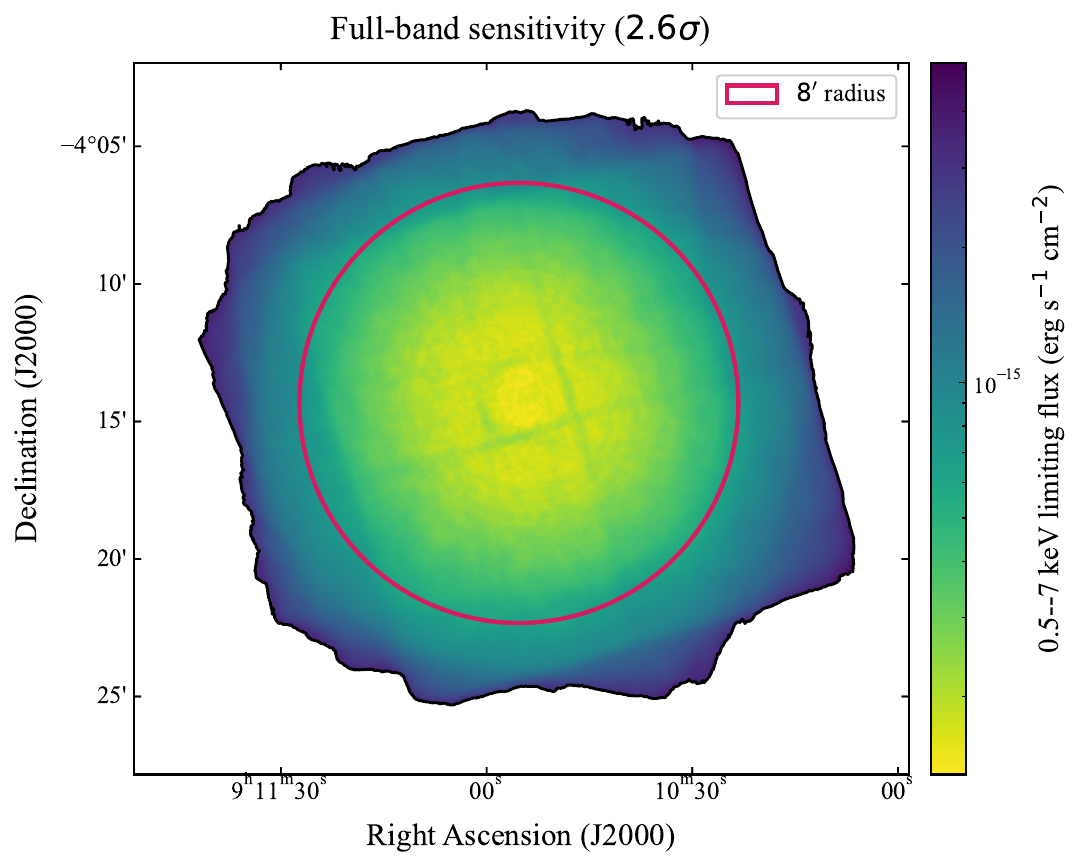}
\includegraphics[width=0.50\textwidth]{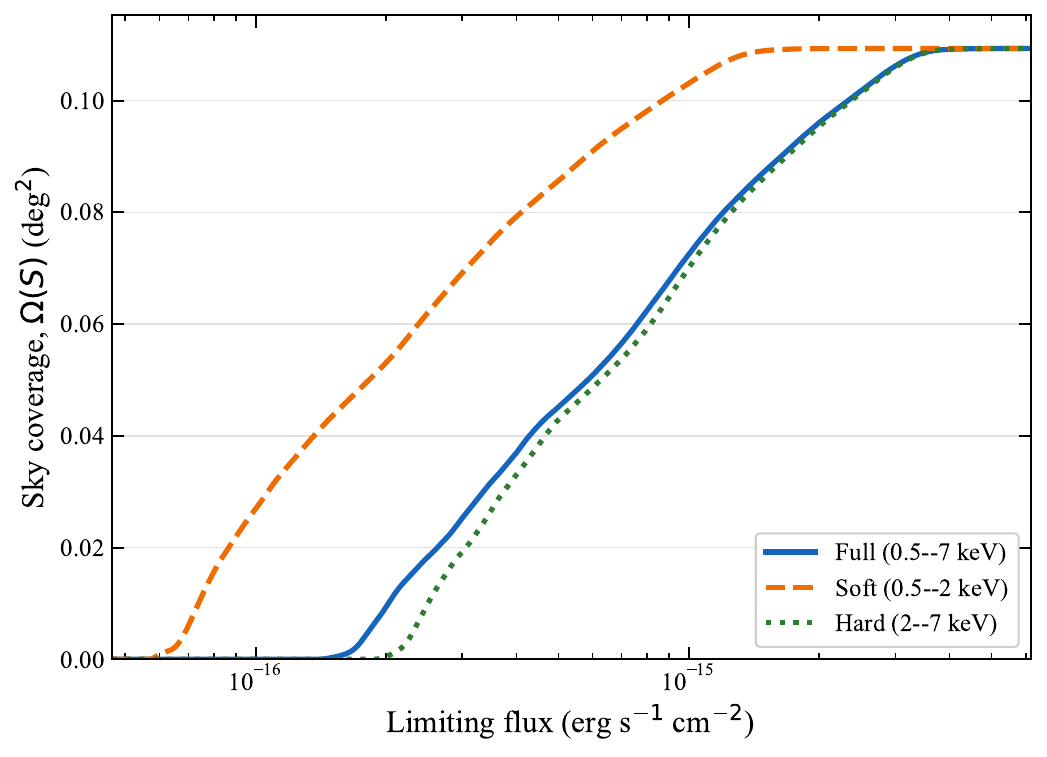}
\caption{{Left:} Full-band sensitivity map. The color bar gives the 0.5--7~keV flux corresponding to the adopted 2.6$\sigma$ \textit{lim\_sens} threshold. The black contour traces the common exposure footprint, and the magenta circle marks the $8\arcmin$-radius region used for the protocluster analysis. {Right:} Sky coverage in the full (blue solid), soft (orange dashed), and hard (green dotted) bands. \label{fig:sensimap}}
\end{figure*}

\section{X-ray properties of protocluster members and field high-redshift galaxies}\label{sec:4}

\subsection{Target sample and individual X-ray analysis}
\label{sec:protocluster_individual}

We compiled a list of targets for X-ray analysis from three source categories:
(1) [\iona{O}{iii}] emitters identified via [\iona{O}{iii}]+H$\beta$ emission from JWST/NIRCam WFSS observations of the J0910$-$0414 field \citep{Wang2026ApJ};
(2) spectroscopically confirmed LAEs and [\iona{C}{ii}] emitters with redshifts from the Deep Extragalactic Imaging Multi-Object Spectrograph (DEIMOS) on Keck, the Inamori-Magellan Areal Camera and Spectrograph (IMACS) on Magellan, or the Atacama Large Millimeter/submillimeter Array (ALMA) \citep{Wang2024ApJ...962L..11W}, all at $6.54 < z < 6.64$ (some of which are also detected as [\iona{O}{iii}] emitters); and
(3) sources with redshifts measured by follow-up JWST/NIRSpec MSA spectroscopy from GO~2028 (PIs: F. Wang and J. Yang), using the PRISM and G395M gratings. These observations yielded redshifts for 23 robust [\iona{O}{iii}]+H$\beta$ emitters at $1.42<z<8.34$, the independently [\iona{C}{ii}]/H$\alpha$-confirmed member SMG-B at $z=6.627$, and one Lyman-break galaxy (LBG) at $z=6.665$. Five of these sources lie within the adopted protocluster interval $6.53<z<6.73$: three [\iona{O}{iii}]+H$\beta$ emitters, SMG-B, and the LBG (Malczewski et al. in prep.).
Sources appearing in multiple categories were cross-matched by position ($< 0\farcs5$) and merged to avoid duplication.
{We restrict the analysis to targets with off-axis angles less than $8\arcmin$, where the exposure is deepest (see Figure~\ref{fig:xexpmap}), the sensitivity is best (see Figure~\ref{fig:sensimap}), and the 90\% encircled-energy radii at the target positions span $R_{90}=1.07\arcsec$--$6.29\arcsec$.} After further excluding targets with overlapping extraction regions (see below), 51 unique targets remain (hereafter referred to as our target sample): 46 [\iona{O}{iii}] emitters, 9 spectroscopically confirmed LAEs/[\iona{C}{ii}] emitters (of which 5 are also [\iona{O}{iii}] emitters), and 1 LBG confirmed by JWST/NIRSpec. Among these, 21 objects have secure spectroscopic redshifts within the adopted protocluster interval $6.53<z<6.73$, including 16 [\iona{O}{iii}] emitters. Centered at $z=6.63$, this interval corresponds to approximately $-3930<\Delta v_{\mathrm{LOS}}<+3930~\mathrm{km~s^{-1}}$, or a total velocity width of $\Delta v_{\mathrm{tot}}\simeq7860~\mathrm{km~s^{-1}}$. It is therefore considerably broader than the typical windows of $|\Delta v_{\mathrm{LOS}}|\lesssim1000$--$2000~\mathrm{km~s^{-1}}$ used in clustering and protocluster-membership analyses. We refer to these 21 objects as protocluster members for brevity, while emphasizing that they constitute a broad redshift-selected sample rather than a strict dynamical membership census. The remaining 30 targets comprise 27 [\iona{O}{iii}] emitters at $z<6.53$ and 3 at $z>6.73$. The central quasar is not included in this target sample and is excluded from all subsequent stacking and AGN incidence calculations, since the protocluster was identified by searching for galaxies around this already known quasar.

\par
We first cross-matched the positions of the 51 retained targets with the main X-ray catalog using a match radius of 3\arcsec, but none were found to have X-ray counterparts. We therefore proceeded to estimate their X-ray photometric properties following Section~2 of \citet{Vito2016MNRAS.463..348V}. We derived the upper limits with the same Bayesian method used for the main catalog. Before extracting the photometry, we masked cataloged X-ray sources and neighboring high-redshift targets to reduce contamination. The main steps are as follows:

 \begin{enumerate}[noitemsep, topsep=1pt]

\item We computed the 90\% encircled-energy radii for the targets ($R_{90\%\mathrm{SRC}}$) and detected X-ray sources ($R^{\mathrm{det}}_{90\%\mathrm{SRC}}$) following Equation~(5) of \citet{Evans2024ApJS..274...22E}.
\item For each target, we defined an annular background-extraction region with inner radius $R_{\mathrm{IN}}=1.1\times R_{90\%\mathrm{SRC}}$ and outer radius $R_{\mathrm{OUT}}=R_{\mathrm{IN}}+20~\mathrm{pixels}$.

\item For each detected source in the main X-ray catalog, we defined a background annulus with inner radius $R^{\mathrm{BKG\_det}}_{\mathrm{IN}}=n\times R^{\mathrm{det}}_{90\%\mathrm{SRC}}$, where $n=1.5$, 1.75, and 2 for sources with $<100$, 100--1000, and $>1000$ net counts, respectively. We manually adjusted $R^{\mathrm{BKG\_det}}_{\mathrm{IN}}$ where necessary to ensure that the emission from each detected source was enclosed within the corresponding circle. The outer radius was set to $R^{\mathrm{BKG\_det}}_{\mathrm{OUT}}=R^{\mathrm{BKG\_det}}_{\mathrm{IN}}+20~\mathrm{pixels}$.

\item For each target, we computed the distances to all other targets and to the detected X-ray sources. If the separation of a pair was smaller than the sum of the inner radii of their respective background regions, we excluded the corresponding target because the overlapping regions could yield unreliable flux estimates. This screening excluded six targets: five because they were too close to another target and one because it lies $2.41\arcsec$ from the cataloged X-ray source XID~30. Because this separation is much larger than the $0.11\arcsec$ positional uncertainty of XID~30, we regard the X-ray source as a nearby contaminant rather than a secure counterpart to the target. The other five excluded targets are also not individually detected in X-rays; therefore, none of the six excluded targets is considered individually X-ray detected.

\item 
To avoid contamination from detected X-ray sources, we masked all sources listed in the main catalog. For each source, the masked region was defined as a circle with a radius of $R^{\mathrm{BKG\_det}}_{\mathrm{IN}}$. The masked areas were refilled with Poisson noise sampled from the background region using the CIAO tool \textit{dmfilth}. We used \textit{roi} to exclude other detected sources from the sampling annuli before refilling the masked pixels.

\item 
For each target, we began with the source-masked image and additionally masked all other targets in the same manner. We used the resulting image for aperture photometry and stacking.

\end{enumerate}
Using the source and background counts obtained from this procedure, we calculated $P_{\rm B}$ for all 51 targets. The minimum $P_{\rm B}$ value among the full, soft, and hard bands ranges from 0.014 to 1.000 for the 51 targets. Thus, none satisfies our adopted detection criterion of $P_{\rm B}<0.01$. We therefore derived their 90\% confidence-level upper limits using the same Bayesian method of \citet{Kraft1991ApJ...374..344K} adopted in Section~\ref{sec:catalog_photometry}. We converted the count-rate limits to full-band flux limits corrected for Galactic absorption using the merged response files and an absorbed power-law model with $\Gamma=1.8$, a commonly adopted value for faint high-redshift AGN and galaxy populations \citep[e.g.,][]{Vito2016MNRAS.463..348V}. {The HI4PI survey gives Galactic column densities of $N_{\rm H}=(2.53$--$2.82)\times10^{20}~\mathrm{cm}^{-2}$ across the 51 target positions \citep{HI4PI2016A&A...594A.116H}; we adopt the representative value $N_{\rm H}=2.7\times10^{20}~\mathrm{cm}^{-2}$ for the flux conversions.} For sources with measured redshifts, upper limits on the rest-frame 2--10 keV luminosity were subsequently calculated using the corresponding $K$-correction. We emphasize that no correction for intrinsic absorption has been applied to these luminosities. The results are summarized in Table~\ref{tbl:uppercatalog}.

\begingroup
\startlongtable
\begin{deluxetable*}{cccccccccccccc}
\tablewidth{0pt}
\tablecaption{X-ray upper-limit constraints for high-$z$ galaxies\label{tbl:uppercatalog}}
\tablehead{
\colhead{ID} & \colhead{R.A.} & \colhead{Decl.} & \colhead{Off-axis} & \colhead{$R_{90}$} & \colhead{Exp.} & \colhead{Type} & \colhead{$z$} & \colhead{$P_{\rm B}$} & \colhead{F} & \colhead{S} & \colhead{H} & \colhead{Flux} & \colhead{$L_{\rm 2-10~keV}$} \\[-5pt]
\colhead{} & \colhead{(deg)} & \colhead{(deg)} & \colhead{(arcmin)} & \colhead{(arcsec)} & \colhead{(s)} & \colhead{} & \colhead{} & \colhead{} & \multicolumn{3}{c}{(counts)} & \colhead{($10^{-16}$)} & \colhead{($10^{43}$)} \\[-5pt]
\colhead{(1)} & \colhead{(2)} & \colhead{(3)} & \colhead{(4)} & \colhead{(5)} & \colhead{(6)} & \colhead{(7)} & \colhead{(8)} & \colhead{(9)} & \colhead{(10)} & \colhead{(11)} & \colhead{(12)} & \colhead{(13)} & \colhead{(14)}
}
\renewcommand{\arraystretch}{1.1}
\setlength{\tabcolsep}{3pt}
\startdata
1 & 137.70052 & -4.27260 & 2.70 & 1.69 & 793227.05 & [O~III] & 6.211 & 0.153 & 7.31 & 2.56 & 8.48 & 2.31 & 5.05 \\
2 & 137.70671 & -4.25720 & 1.80 & 1.28 & 748059.33 & [O~III] & 5.924 & 0.767 & 3.32 & 2.56 & 3.47 & 1.11 & 2.20 \\
3 & 137.70702 & -4.27966 & 2.82 & 1.75 & 751993.94 & [O~III] & 6.090 & 0.014 & 12.37 & 3.96 & 11.81 & 4.11 & 8.63 \\
{4}\tablenotemark{a} & 137.71552 & -4.22928 & 1.06 & 1.10 & 849616.32 & [O~III] & 6.629 & 0.691 & 4.44 & 2.56 & 4.76 & 1.31 & 3.29 \\
{5}\tablenotemark{a} & 137.71622 & -4.28436 & 2.86 & 1.78 & 779749.62 & [O~III]+LAE & 6.537 & 0.268 & 7.45 & 4.95 & 5.90 & 2.39 & 5.83 \\
{6}\tablenotemark{a} & 137.71680 & -4.21461 & 1.66 & 1.24 & 827669.16 & [O~III] & 6.634 & 0.797 & 3.97 & 2.56 & 4.24 & 1.20 & 3.02 \\
7 & 137.71712 & -4.23127 & 0.91 & 1.09 & 852782.96 & [O~III] & 5.433 & 0.343 & 4.43 & 2.56 & 4.77 & 1.30 & 2.13 \\
8 & 137.72063 & -4.26515 & 1.69 & 1.24 & 761801.80 & [O~III] & 6.188 & 0.530 & 3.25 & 3.75 & 2.56 & 1.07 & 2.32 \\
{9}\tablenotemark{a} & 137.72105 & -4.22559 & 0.97 & 1.09 & 850523.52 & [O~III] & 6.635 & 0.712 & 4.02 & 2.56 & 4.28 & 1.18 & 2.98 \\
10 & 137.72119 & -4.24520 & 0.67 & 1.07 & 864709.12 & [O~III] & 6.382 & 0.477 & 4.49 & 2.56 & 4.75 & 1.30 & 3.02 \\
{11}\tablenotemark{a} & 137.72362 & -4.22296 & 1.03 & 1.10 & 848348.95 & [O~III]+LAE & 6.632 & 1.000 & 3.06 & 2.56 & 3.14 & 0.90 & 2.27 \\
12 & 137.72418 & -4.26185 & 1.43 & 1.17 & 668148.95 & [O~III] & 5.827 & 1.000 & 3.08 & 2.56 & 3.19 & 1.15 & 2.20 \\
13 & 137.72418 & -4.22788 & 0.75 & 1.08 & 861080.95 & [O~III] & 5.515 & 0.259 & 6.37 & 4.13 & 5.24 & 1.85 & 3.13 \\
{14}\tablenotemark{a} & 137.72644 & -4.24801 & 0.60 & 1.07 & 865211.42 & [O~III]+LAE & 6.637 & 0.401 & 4.91 & 4.43 & 3.55 & 1.42 & 3.58 \\
{15}\tablenotemark{a} & 137.72726 & -4.23208 & 0.44 & 1.07 & 864903.85 & [O~III]+[C~II] & 6.627 & 1.000 & 3.34 & 2.56 & 3.49 & 0.97 & 2.43 \\
{16}\tablenotemark{a} & 137.72810 & -4.23555 & 0.24 & 1.08 & 865230.81 & [O~III] & 6.656 & 1.000 & 2.95 & 3.26 & 2.56 & 0.85 & 2.16 \\
17 & 137.72914 & -4.23303 & 0.35 & 1.08 & 865308.17 & [O~III] & 5.705 & 0.573 & 4.37 & 2.60 & 4.66 & 1.26 & 2.30 \\
{18}\tablenotemark{a} & 137.73102 & -4.23663 & 0.14 & 1.10 & 865197.20 & [O~III] & 6.631 & 0.776 & 3.82 & 3.22 & 3.46 & 1.10 & 2.78 \\
19 & 137.73155 & -4.25807 & 1.16 & 1.12 & 583880.79 & [O~III] & 6.384 & 0.302 & 4.29 & 4.00 & 3.30 & 1.84 & 4.27 \\
20 & 137.73338 & -4.24018 & 0.20 & 1.09 & 862895.78 & [O~III] & 5.705 & 0.299 & 5.63 & 4.42 & 4.29 & 1.63 & 2.98 \\
{21}\tablenotemark{a} & 137.73575 & -4.22681 & 0.79 & 1.08 & 853748.62 & [O~III] & 6.640 & 0.735 & 3.98 & 3.33 & 3.51 & 1.16 & 2.94 \\
22 & 137.73647 & -4.24846 & 0.68 & 1.07 & 849495.38 & [O~III] & 6.387 & 0.838 & 3.51 & 3.55 & 2.89 & 1.03 & 2.40 \\
{23}\tablenotemark{a} & 137.73894 & -4.24618 & 0.68 & 1.07 & 860519.12 & [O~III] & 6.633 & 0.751 & 3.48 & 2.56 & 3.69 & 1.01 & 2.55 \\
{24}\tablenotemark{a} & 137.74239 & -4.22685 & 1.01 & 1.10 & 821233.12 & [O~III] & 6.565 & 0.764 & 4.26 & 2.69 & 4.48 & 1.30 & 3.20 \\
25 & 137.74460 & -4.22829 & 1.06 & 1.10 & 791031.45 & [O~III] & 5.745 & 1.000 & 2.56 & 2.56 & 2.56 & 0.81 & 1.50 \\
{26}\tablenotemark{a} & 137.74831 & -4.22918 & 1.22 & 1.13 & 748578.89 & [O~III] & 6.636 & 0.359 & 4.50 & 2.56 & 4.81 & 1.50 & 3.80 \\
{27}\tablenotemark{a} & 137.68238 & -4.13214 & 7.01 & 6.29 & 802070.81 & LAE & 6.572 & 0.288 & 17.26 & 6.75 & 16.97 & 5.38 & 13.30 \\
{28}\tablenotemark{a} & 137.67868 & -4.29297 & 4.49 & 3.08 & 744140.37 & LAE & 6.626 & 0.051 & 14.40 & 4.83 & 14.13 & 4.84 & 12.17 \\
{29}\tablenotemark{a} & 137.69240 & -4.29659 & 4.15 & 2.75 & 825601.41 & LAE & 6.634 & 0.316 & 8.82 & 3.93 & 8.92 & 2.67 & 6.74 \\
30 & 137.70918 & -4.28492 & 3.04 & 1.89 & 759911.35 & [O~III] & 7.970 & 0.468 & 5.75 & 4.55 & 4.69 & 1.89 & 7.05 \\
31 & 137.70724 & -4.27529 & 2.59 & 1.62 & 733450.41 & [O~III] & 4.421 & 0.617 & 4.07 & 2.56 & 4.35 & 1.39 & 1.45 \\
32 & 137.72516 & -4.22530 & 0.87 & 1.08 & 856992.18 & [O~III] & 6.200 & 0.315 & 5.46 & 3.98 & 4.45 & 1.59 & 3.48 \\
33 & 137.69984 & -4.27638 & 2.90 & 1.80 & 812162.83 & [O~III] & 4.669 & 0.201 & 6.38 & 2.56 & 7.11 & 1.96 & 2.32 \\
34 & 137.73221 & -4.23139 & 0.46 & 1.07 & 865077.03 & [O~III] & 4.877 & 1.000 & 3.18 & 3.60 & 2.59 & 0.92 & 1.19 \\
35 & 137.69717 & -4.23891 & 1.99 & 1.35 & 643195.18 & [O~III] & 5.060 & 0.780 & 3.85 & 2.56 & 4.07 & 1.50 & 2.10 \\
36 & 137.72197 & -4.26105 & 1.43 & 1.17 & 675720.81 & [O~III] & 5.313 & 0.067 & 5.96 & 5.52 & 3.63 & 2.21 & 3.45 \\
37 & 137.71991 & -4.21468 & 1.58 & 1.21 & 822917.23 & [O~III] & 5.433 & 0.385 & 4.26 & 3.91 & 3.44 & 1.30 & 2.12 \\
38 & 137.71880 & -4.23399 & 0.75 & 1.08 & 859225.16 & [O~III] & 5.435 & 0.553 & 4.82 & 3.56 & 4.21 & 1.40 & 2.30 \\
39 & 137.72298 & -4.24237 & 0.49 & 1.07 & 863879.49 & [O~III] & 5.442 & 1.000 & 3.12 & 2.56 & 3.23 & 0.90 & 1.49 \\
40 & 137.70969 & -4.28993 & 3.31 & 2.07 & 762151.01 & [O~III] & 5.586 & 0.747 & 4.28 & 3.68 & 3.87 & 1.40 & 2.45 \\
41 & 137.71545 & -4.24013 & 0.90 & 1.09 & 855093.52 & [O~III] & 6.216 & 0.438 & 5.46 & 2.56 & 5.89 & 1.60 & 3.50 \\
42 & 137.72104 & -4.22022 & 1.25 & 1.13 & 858789.32 & [O~III] & 6.225 & 0.385 & 5.35 & 5.24 & 3.46 & 1.56 & 3.43 \\
43 & 137.73274 & -4.27358 & 2.09 & 1.39 & 815501.87 & [O~III] & 6.396 & 0.654 & 3.85 & 2.56 & 4.20 & 1.18 & 2.76 \\
44 & 137.70422 & -4.24752 & 1.65 & 1.23 & 685725.77 & [O~III] & 6.440 & 1.000 & 2.56 & 2.56 & 2.56 & 0.93 & 2.21 \\
{45}\tablenotemark{a} & 137.75539 & -4.25627 & 1.83 & 1.29 & 780907.35 & [O~III]+LAE & 6.622 & 0.805 & 3.88 & 2.59 & 4.12 & 1.24 & 3.12 \\
{46}\tablenotemark{a} & 137.72334 & -4.23589 & 0.46 & 1.07 & 865202.71 & [C~II] & 6.627 & 0.744 & 3.54 & 2.69 & 3.62 & 1.02 & 2.58 \\
{47}\tablenotemark{a} & 137.71886 & -4.22457 & 1.10 & 1.11 & 843949.81 & [O~III] & 6.635 & 0.695 & 3.85 & 3.10 & 3.57 & 1.14 & 2.88 \\
48 & 137.70807 & -4.27140 & 2.37 & 1.51 & 711776.28 & [O~III] & 7.498 & 1.000 & 2.58 & 2.56 & 2.58 & 0.91 & 2.97 \\
49 & 137.71240 & -4.27505 & 2.43 & 1.54 & 745459.27 & [O~III] & 8.338 & 0.546 & 4.93 & 3.69 & 4.32 & 1.65 & 6.77 \\
{50}\tablenotemark{a} & 137.73077 & -4.27184 & 1.98 & 1.34 & 777855.95 & LBG & 6.665 & 0.565 & 5.00 & 3.91 & 4.24 & 1.61 & 4.09 \\
{51}\tablenotemark{a} & 137.71713 & -4.22586 & 1.11 & 1.11 & 852180.09 & [O~III] & 6.632 & 0.714 & 4.43 & 2.56 & 4.74 & 1.30 & 3.28 \\
\enddata
\tablenotetext{a}{{Confirmed member of the adopted protocluster sample at $6.53<z<6.73$.}}
\tablecomments{Columns are as follows: (1) Source ID; (2)--(3) Source coordinates (J2000); (4) Off-axis angle in units of arcminutes; (5) Radius enclosing 90\% of the encircled energy fraction in units of arcseconds; (6) Exposure time on the source region in units of seconds; (7) Source type: [O~III] denotes [\iona{O}{iii}] emitters spectroscopically identified from JWST/NIRCam WFSS and/or JWST/NIRSpec MSA observations. Composite labels indicate sources independently confirmed as LAEs or [\iona{C}{ii}] emitters. LAE denotes spectroscopically confirmed Ly$\alpha$ emitters, [C~II] denotes [\iona{C}{ii}] emitters identified by ALMA, and LBG denotes Lyman-break galaxies confirmed by JWST/NIRSpec; (8) Spectroscopic redshift; (9) Minimum $P_{\rm B}$ value among the full, soft, and hard bands; (10)--(12) 90\% confidence-level upper limits on the source counts in the full, soft, and hard bands, respectively; (13) 90\% confidence-level upper limit on the full-band flux (in units of $10^{-16}~\mathrm{erg\,s^{-1}\,cm^{-2}}$); (14) 90\% confidence-level upper limit on the rest-frame 2--10~keV luminosity (in units of $10^{43}~\mathrm{erg\,s^{-1}}$), calculated assuming $\Gamma=1.8$.}
\end{deluxetable*}
\endgroup

\subsection{Stacking analysis} \label{sec:stack}
To probe fainter, sub-threshold X-ray emission, we performed stacking analyses for several subclasses of our sample, using the photometric measurements described in Section~\ref{sec:protocluster_individual}.
The stacked samples include: (i) all 46 [\iona{O}{iii}] emitters spanning $4.42 < z < 8.34$ (mean $\bar{z}=6.20$); (ii) the 16 [\iona{O}{iii}] emitters at $6.53 < z < 6.73$ that are protocluster members (mean $\bar{z}=6.62$); (iii) the 7 spectroscopically confirmed LAEs from \citet{Wang2024ApJ...962L..11W}, of which 4 are also [\iona{O}{iii}] emitters included in (ii) (mean $\bar{z}=6.61$); and (iv) all 21 protocluster members as defined in Section~\ref{sec:protocluster_individual} (mean $\bar{z}=6.62$).

For each quantitative stack, we summed the counts measured within the $R_{90\%\mathrm{SRC}}$ apertures and the local background counts after scaling them to the corresponding source-aperture areas. We calculated $P_{\rm B}$ and, following \citet{Kraft1991ApJ...374..344K}, derived the 90\% upper limit on the combined net source counts. After applying the 90\% EEF aperture correction, we divided the count upper limit by the summed effective exposure, so that the inferred flux and luminosity limits represent exposure-weighted sample averages rather than total emission from the subclass. The count-rate limits were converted to Galactic-absorption-corrected fluxes using the $\Gamma=1.8$ response conversion described in Section~\ref{sec:protocluster_individual}, and then to rest-frame 2--10~keV luminosities at the mean redshift of each subclass; no intrinsic-absorption correction was applied.

For the four stacked subsamples, the minimum $P_{\rm B}$ value among the full, soft, and hard bands ranges from 0.060 to 0.606. Therefore, none of the stacked samples satisfies our adopted detection criterion of $P_{\rm B}<0.01$; the corresponding 90\% confidence-level upper limits, calculated assuming $\Gamma=1.8$, are summarized in Table~\ref{tbl:stacking}. For all 21 protocluster members, the total effective exposure time reaches $\sim17$~Ms; this non-detection implies an average X-ray luminosity of $\leq9.54\times10^{42}~\mathrm{erg~s^{-1}}$. For reference, adopting an X-ray bolometric correction of $k_{\rm bol}=25$ and a radiative efficiency of $\epsilon=0.1$ \citep[see, e.g.,][]{Lusso2012MNRAS}, this limit corresponds to a sample-averaged black-hole accretion rate of $\dot{M}_{\rm BH}<0.042~M_\odot\,\mathrm{yr}^{-1}$ and an Eddington ratio of $\lambda_{\rm Edd}<0.19\,(M_{\rm BH}/10^7M_\odot)^{-1}$. For the 16 [\iona{O}{iii}] members, the corresponding values are $\dot{M}_{\rm BH}<0.020~M_\odot\,\mathrm{yr}^{-1}$ and $\lambda_{\rm Edd}<0.090\,(M_{\rm BH}/10^7M_\odot)^{-1}$. Because the black-hole masses are not measured and the conversion assumes that the intrinsic X-ray emission is not strongly suppressed, these accretion-rate and Eddington-ratio limits should be treated only as illustrative. We find no clear evidence for widespread X-ray-bright SMBH accretion among the protocluster members. However, the stacked limit constrains the sample-averaged luminosity rather than providing a uniform detection threshold for every galaxy. It therefore does not exclude moderate-luminosity AGN hosted by a subset of the members, particularly if their X-ray emission is obscured or intrinsically weak (see Section~\ref{sec:nondetection_discussion} for further discussion).

\begin{figure}[htbp]
  \centering
  \includegraphics[width= 3.3in]{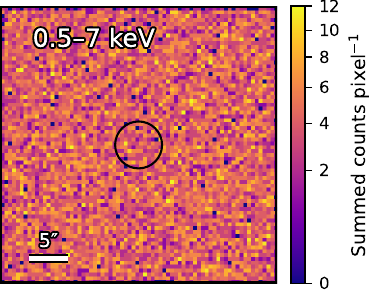}
    \caption{Summed 35\arcsec$\times$35\arcsec full-band image of the 38 targets with off-axis angles smaller than 2\arcmin. The circle at the center has a radius of 3\arcsec. This image is shown for visualization and is not used to derive the quantitative stacking limits in Table~\ref{tbl:stacking}. \label{fig:imstack}}
  \end{figure}

For Figure~\ref{fig:imstack}, we also constructed a visualization-only full-band image stack. We selected the 38 targets with off-axis angles smaller than $2\arcmin$, where the \chandra\ PSF varies only mildly, and summed their count-image cutouts using \textit{dmimgcalc}. This image stack is not used to derive the quantitative limits in Table~\ref{tbl:stacking}.

\begin{deluxetable*}{lccccccc}[t!]
\tablewidth{0pt}
\tabletypesize{\scriptsize}
\tablecaption{Stacking analysis of subclasses of our targets\label{tbl:stacking}}
\tablehead{
\colhead{Subclasses} & \colhead{Num} & \colhead{Total Exp.} & \colhead{$z_{\rm mean}$} & \colhead{F} & \colhead{S} & \colhead{H} & \colhead{$L_{\rm 2-10~keV}$}\\[-5pt]
\colhead{} & \colhead{} & \colhead{(Ms)} & \colhead{} & \multicolumn{3}{c}{(counts)} & \colhead{($10^{42}~\mathrm{erg\,s^{-1}}$)}\\[-5pt]
\colhead{(1)} & \colhead{(2)} & \colhead{(3)} & \colhead{(4)} & \colhead{(5)} & \colhead{(6)} & \colhead{(7)} & \colhead{(8)}
}
\startdata
All [O~III] emitters & 46 & 37.0 & 6.20 & 17.41 & 7.90 & 18.41 & 2.57  \\
{[O~III] members} & 16 & 13.4 & 6.62 & 9.68 & 4.65 & 10.59 & 4.55  \\
All LAEs & 7 & 5.6 & 6.61 & 28.59 & 8.64 & 29.18 & 31.67  \\
All members & 21 & 17.4 & 6.62 & 26.38 & 7.48 & 29.88 & 9.54  \\
\enddata
\tablecomments{Columns are as follows: (1) Subclass name; (2) Number of sources; (3) Total effective exposure time in units of Ms; (4) Mean redshift of the subclass; (5)--(7) aperture-corrected 90\% confidence-level upper limits on the stacked source counts in the full (0.5--7~keV), soft (0.5--2~keV), and hard (2--7~keV) bands, respectively; (8) 90\% confidence-level upper limit on the sample-averaged rest-frame 2--10~keV luminosity (in units of $10^{42}~\mathrm{erg\,s^{-1}}$), computed from the full-band flux upper limit assuming $\Gamma=1.8$.}
\end{deluxetable*}

\subsection{The properties and expected X-ray emissions of [O~III] emitters}
\label{sec:host_expected_xray}
We examine the properties of the 16 [\iona{O}{iii}] emitters in the adopted protocluster interval $6.53<z<6.73$. We designate these sources O3-01--O3-16 in order of increasing R.A. Their coordinates, spectroscopic redshifts, data coverage, and inferred properties are listed in Table~\ref{tbl:host_oiii}.

Fifteen members are covered by the JWST/NIRCam imaging, for which we use the F115W, F200W, and F356W total fluxes. O3-16 lies outside the available NIRCam footprint and is fitted with HST/WFC3 F125W and F160W photometry. Following the use of a minimum fractional photometric error in comparable ASPIRE SED fitting \citep[5\%;][]{Champagne2025ApJII}, we adopt a more conservative 10\% uncertainty floor and add it in quadrature to the cataloged uncertainty of each photometric point. We fit the broad-band photometry with {\sc bagpipes} \citep{Carnall2018MNRAS}, fixing each source at its spectroscopic redshift and including the independently measured [\iona{O}{iii}] $\lambda5007$ flux in the likelihood\citet{Champagne2025ApJII}.

Our primary model adopts a delayed-exponential SFH and the 2016 version of the BC03 stellar-population models \citep{Bruzual2003MNRAS}, updated with the MILES stellar library \citep{FalconBarroso2011A&A} and constructed with the Kroupa \& Boily IMF \citep{Kroupa2002MNRAS}, as implemented in {\sc bagpipes} \citep{Carnall2018MNRAS}. The model assumptions and priors are summarized in Table~\ref{tbl:sed_setup}. Fits were performed with the \texttt{nautilus} sampler using 200 live points, and the reported SFR is averaged over the most recent 100~Myr. The fits contain stellar and nebular emission but no separate AGN continuum component; the inferred galaxy properties therefore adopt a galaxy-only interpretation.

\begin{deluxetable*}{lll}
\tablewidth{0pt}
\tabletypesize{\scriptsize}
\setlength{\tabcolsep}{5pt}
\tablecaption{Primary SED-fitting model and priors\label{tbl:sed_setup}}
\tablehead{
\colhead{Component} & \colhead{Adopted treatment} & \colhead{Prior or setting}
}
\startdata
Redshift & Spectroscopic redshift & Fixed for each source \\
Stellar population & BC03/MILES; Kroupa IMF & Fixed \\
Star-formation history & Delayed exponential, ${\rm SFR}(t)\propto t\exp(-t/\tau)$ & Primary model \\
Stellar-population age & 1~Myr to the age of the Universe at $z_{\rm spec}$ & Uniform \\
$\tau$ & 0.01--2~Gyr & Log-uniform \\
Formed stellar mass & $6<\log(M_{\rm formed}/M_\odot)<12$ & Uniform in $\log M_{\rm formed}$ \\
Stellar metallicity & $0.01<Z/Z_\odot<1$ & Log-uniform \\
Dust attenuation & Calzetti law; $0<A_V<3$ & Uniform in $A_V$ \\
Nebular emission & $-4<\log U<-1$ & Uniform in $\log U$ \\
Birth-cloud lifetime & 10~Myr & Fixed \\
\enddata
\tablecomments{The delayed-$\tau$ model provides the primary mass and SFR estimates. Constant and five-bin non-parametric continuity SFHs are fitted separately to assess SFH-dependent systematics.}
\end{deluxetable*}

We repeat every fit with a constant SFH and a five-bin non-parametric continuity SFH. 
Because broad-band photometry often cannot discriminate among plausible SFH priors \citep{Carnall2019ApJ}, we use the delayed-$\tau$ 16th--84th percentiles as the within-model statistical uncertainties and report the range of the three posterior medians separately.

The primary fits span $8.45<\log(M_\star/M_\odot)<9.91$ and $1.6<\mathrm{SFR}/(M_\odot\,\mathrm{yr}^{-1})<28.6$, with sample medians of 9.05 and $7.1~M_\odot\,\mathrm{yr}^{-1}$, respectively. The inferred properties remain sensitive to the assumed SFH: the median source-by-source span among the three SFH models is 0.53~dex in stellar mass and 0.20~dex in logarithmic SFR. The two-band fit for O3-16 does not sample the rest-frame optical continuum and is consequently less well constrained than the three-band NIRCam fits.

For 15 members, the [\iona{O}{iii}] $\lambda5007$ flux is measured from the NIRCam/WFSS two-dimensional spectra, including two lower-S/N lines extracted at the known spectroscopic redshifts. The O3-16 measurement is obtained from its NIRSpec/G395M line profile. The WFSS measurements follow the ASPIRE line-extraction and calibration procedures \citep{Wang2023ApJ,Wang2026ApJ}. To allow for residual differences among measurement routes and line quality, we conservatively add systematic uncertainty floors of 20\% for the 13 primary WFSS measurements, 25\% for the lower-S/N O3-04 and O3-05 WFSS measurements, and 30\% for O3-16. For O3-16, the adopted floor also allows for NIRSpec/MSA aperture and path-loss uncertainties \citep{Ferruit2022AandA}. These systematic terms are added in quadrature to the statistical line-flux errors. We adopt $F_{5007}/F_{4959}=2.98$ for the doublet \citep{Storey2000MNRAS}. The resulting observed [\iona{O}{iii}] luminosities, not corrected for dust attenuation within the host galaxies, span $41.575<\log(L_{5007}/\mathrm{erg\,s^{-1}})<43.047$.

For the 15 members with F356W photometry, the measured [\iona{O}{iii}] doublet contributes 12.3--58.4\% of the total F356W flux density, with a median of 31.5\%. For every source, the measured line contribution is smaller than the total F356W flux. The remaining broad-band flux includes the stellar continuum, H$\beta$, and weaker emission lines.

{We estimate the rest-frame 2--10~keV luminosity from low-mass and high-mass X-ray binaries (LMXBs and HMXBs) using Equation~15 of \citet{Lehmer2016ApJ}, which includes a stellar-mass-dependent LMXB component and an SFR-dependent HMXB component, both evolving with redshift. Because this relation was constrained primarily at $z\lesssim2.5$, its application at $z\simeq6.6$ is an extrapolation. Moreover, the HMXB luminosity per unit SFR increases toward lower metallicity \citep{Lehmer2024ApJ}. The metallicities of the individual galaxies are not sufficiently constrained to apply such a correction, and we therefore treat the following values as nominal estimates that may underestimate the HMXB contribution.} The individual posterior medians span $4.9\times10^{40}$--$9.0\times10^{41}~\mathrm{erg\,s^{-1}}$ and lie factors of 33--583 below the corresponding individual X-ray upper limits. Weighting the medians by the same effective exposures used for the 16-object [\iona{O}{iii}]-member stack gives $\langle L_{2-10}^{\rm XRB}\rangle_{\rm exp}=2.77\times10^{41}~\mathrm{erg\,s^{-1}}$, approximately 16 times below the 90\% upper limit of $4.55\times10^{42}~\mathrm{erg\,s^{-1}}$ from the 16-source [\iona{O}{iii}]-member stack. {Even an increase approaching one order of magnitude, as allowed by the low-metallicity results of \citet{Lehmer2024ApJ}, would not exceed the stacking upper limit. The non-detections therefore do not constrain the expected XRB emission.}

We separately estimate the rest-frame 0.5--2~keV emission associated with the hot interstellar medium from the SFR scaling of \citet{Mineo2012MNRAS.426.1870M}. The individual posterior medians span $1.4\times10^{39}$--$2.4\times10^{40}~\mathrm{erg\,s^{-1}}$, and their exposure-weighted mean is $7.6\times10^{39}~\mathrm{erg\,s^{-1}}$. These predicted luminosities are far below our X-ray upper limits, so the non-detections do not constrain this component. We retain the estimate in its calibrated soft band rather than adding it to the 2--10~keV XRB prediction.

As an extreme AGN-only test, we assign all of the observed [\iona{O}{iii}] emission to an AGN and apply the observed-luminosity [\iona{O}{iii}]--X-ray relation of \citet{Ueda2015ApJ}. The inferred luminosities, $L_{2-10}\simeq(0.51$--$9.0)\times10^{44}~\mathrm{erg\,s^{-1}}$, exceed the individual X-ray upper limit for every source. The data therefore disfavor a hypothesis in which the line emission of all 16 galaxies is dominated by unobscured, X-ray-normal AGN. This result applies only to the extreme hypothesis for the full sample and does not rule out AGN in individual galaxies. For an individual source, star formation may contribute substantially to the observed [\iona{O}{iii}] emission, and the relation has a dispersion of approximately 0.63~dex \citep{Berney2015MNRAS}. Moreover, strongly obscured or intrinsically X-ray-weak AGN may remain undetected.

An independent limiting-case test follows from the rest-frame 2500~\AA\ continuum and the $\alpha_{\rm ox}$--$L_{2500}$ relation of \citet{Nanni2017A&A}. We estimate $L_{2500}$ from F200W for 15 sources and from F160W for O3-16, adopting the unobscured-quasar continuum $f_\nu\propto\nu^{-0.5}$ \citep{Richards2001AJ}, consistent with the convention of \citet{Nanni2017A&A}. We convert the inferred monochromatic 2~keV luminosity to rest-frame 2--10~keV assuming $\Gamma=1.8$. Photometric uncertainties and an intrinsic $\alpha_{\rm ox}$ scatter of 0.12 are propagated. If the total measured UV continuum were produced by an unobscured, radio-quiet type-1 AGN, the inferred range would be $42.74<\log(L_{2-10}/\mathrm{erg\,s^{-1}})<43.60$. The exposure-weighted mean, $2.02\times10^{43}~\mathrm{erg\,s^{-1}}$, is 4.45 times the 16-object stacking limit, although only two individual median predictions exceed their corresponding limits. The stacked comparison disfavors a scenario in which the UV continua of all 16 sources are dominated by unobscured type-1 AGN, but it does not exclude weaker or obscured accretion in individual galaxies because starlight may dominate the measured UV continuum.

\begin{figure*}[htbp]
  \centering
  \includegraphics[width=\textwidth]{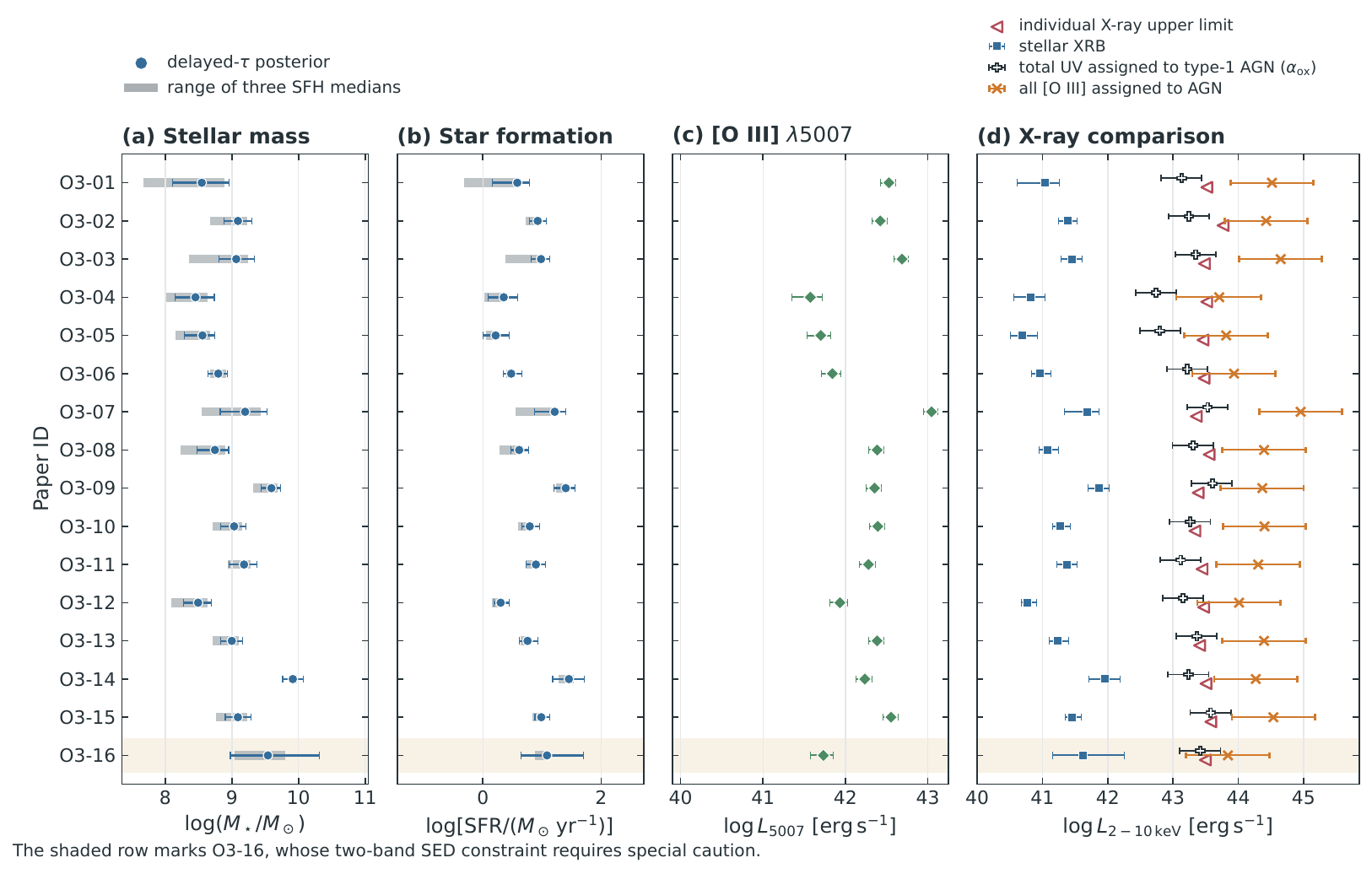}
  \caption{SED, [\iona{O}{iii}], and X-ray diagnostics for the 16 [\iona{O}{iii}] members in the host-property sample. Panels (a) and (b) show the delayed-$\tau$ posterior medians and 16th--84th percentile intervals for stellar mass and SFR; thick gray bars show the range of posterior medians across the delayed-$\tau$, constant, and continuity SFHs. Panel (c) shows the observed $L_{5007}$ measurements, which are not corrected for dust attenuation within the host galaxies; their intervals combine statistical and adopted systematic line-flux errors. Panel (d) compares the extrapolated stellar 2--10~keV XRB prediction, the individual 90\% X-ray upper limit, and two AGN-only counterfactuals: the X-ray luminosity obtained by assigning all measured [\iona{O}{iii}] emission to an AGN, and the $\alpha_{\rm ox}$ prediction obtained by assigning the total UV continuum to an unobscured type-1 AGN. The XRB intervals propagate the delayed-$\tau$ SED posterior. The orange intervals combine the $L_{5007}$ measurement errors with the 0.63~dex intrinsic dispersion of the local [\iona{O}{iii}]--X-ray relation, while the $\alpha_{\rm ox}$ intervals combine photometric errors with an intrinsic $\alpha_{\rm ox}$ scatter of 0.12. Left-pointing open symbols denote X-ray upper limits. The shaded row identifies O3-16, whose two-band SED constraint requires special caution. \label{fig:host_oiii}}
\end{figure*}

The [\iona{O}{iii}]-based values exceed the $\alpha_{\rm ox}$ estimates by 0.42--1.42~dex, with a median offset of 1.03~dex. This difference reflects the distinct observables and calibration samples: the [\iona{O}{iii}] relation is based on narrow-line emission from local hard-X-ray-selected AGN \citep{Ueda2015ApJ}, whereas the $\alpha_{\rm ox}$ relation describes luminous unobscured quasars \citep{Nanni2017A&A}. However, in our high redshift galaxies sample, young stellar populations can power strong [\iona{O}{iii}] emission \citep{Matthee2023ApJ}, while stars can also dominate the rest-frame UV continuum. Figure~\ref{fig:host_oiii} summarizes the stellar-mass, SFR, [\iona{O}{iii}], and X-ray comparisons for the 16 sources.

\begin{deluxetable*}{lrrrlccccccc}
\setlength{\tabcolsep}{1pt}
\tablewidth{0pt}
\tabletypesize{\scriptsize}
\tablecaption{Properties of the 16 [O III] protocluster members\label{tbl:host_oiii}}
\tablehead{
\colhead{ID} & \colhead{R.A.} & \colhead{Decl.} & \colhead{$z_{\rm spec}$} & \colhead{Spec.} & \colhead{$\log(M_\star/M_\odot)$} & \colhead{SFR} & \colhead{$\log L_{5007}$} & \colhead{$\log L_{\rm X}^{\rm XRB}$} & \colhead{$\log L_{\rm X}^{\rm gas}$} & \colhead{$\log L_{\rm X}^{\rm OIII}$} & \colhead{$\log L_{\rm X}^{\alpha_{\rm ox}}$}\\[-5pt]
\colhead{} & \multicolumn{2}{c}{(deg; J2000)} & \colhead{} & \colhead{} & \colhead{} & \colhead{($M_\odot\,\mathrm{yr}^{-1}$)} & \multicolumn{5}{c}{($\mathrm{erg\,s^{-1}}$)}
}
\startdata
O3-01 & 137.71552 & $-4.22928$ & 6.6290 & W+P+G & $8.55^{+0.41}_{-0.44}$ & $3.8^{+2.4}_{-2.4}$ & $42.53^{+0.08}_{-0.10}$ & $41.05^{+0.22}_{-0.43}$ & $39.50^{+0.21}_{-0.43}$ & $44.52^{+0.63}_{-0.64}$ & $43.13^{+0.31}_{-0.32}$ \\
O3-02\tablenotemark{a} & 137.71622 & $-4.28436$ & 6.5370 & W+P & $9.09^{+0.21}_{-0.20}$ & $8.5^{+3.4}_{-2.4}$ & $42.42^{+0.08}_{-0.10}$ & $41.39^{+0.14}_{-0.14}$ & $39.85^{+0.15}_{-0.14}$ & $44.43^{+0.63}_{-0.64}$ & $43.24^{+0.31}_{-0.31}$ \\
O3-03 & 137.71680 & $-4.21461$ & 6.6340 & W+P+G & $9.07^{+0.27}_{-0.26}$ & $9.6^{+4.0}_{-3.0}$ & $42.69^{+0.08}_{-0.10}$ & $41.45^{+0.15}_{-0.16}$ & $39.90^{+0.15}_{-0.16}$ & $44.65^{+0.63}_{-0.64}$ & $43.34^{+0.31}_{-0.31}$ \\
O3-04 & 137.71713 & $-4.22586$ & 6.6322 & W+P & $8.45^{+0.28}_{-0.31}$ & $2.3^{+1.6}_{-1.0}$ & $41.57^{+0.15}_{-0.23}$ & $40.82^{+0.22}_{-0.27}$ & $39.27^{+0.23}_{-0.27}$ & $43.71^{+0.64}_{-0.66}$ & $42.74^{+0.31}_{-0.31}$ \\
O3-05 & 137.71886 & $-4.22457$ & 6.6347 & W+G & $8.56^{+0.19}_{-0.27}$ & $1.6^{+1.1}_{-0.6}$ & $41.70^{+0.12}_{-0.17}$ & $40.69^{+0.22}_{-0.19}$ & $39.14^{+0.22}_{-0.21}$ & $43.81^{+0.64}_{-0.65}$ & $42.80^{+0.31}_{-0.31}$ \\
O3-06 & 137.72105 & $-4.22559$ & 6.6350 & W+P+G & $8.80^{+0.14}_{-0.15}$ & $3.0^{+1.6}_{-0.8}$ & $41.84^{+0.10}_{-0.13}$ & $40.97^{+0.17}_{-0.13}$ & $39.40^{+0.19}_{-0.14}$ & $43.93^{+0.64}_{-0.64}$ & $43.22^{+0.31}_{-0.31}$ \\
O3-07\tablenotemark{a} & 137.72362 & $-4.22296$ & 6.6320 & W+G & $9.20^{+0.33}_{-0.37}$ & $16.4^{+9.2}_{-9.0}$ & $43.05^{+0.08}_{-0.10}$ & $41.68^{+0.19}_{-0.35}$ & $40.13^{+0.19}_{-0.34}$ & $44.95^{+0.63}_{-0.64}$ & $43.53^{+0.31}_{-0.31}$ \\
O3-08\tablenotemark{a} & 137.72644 & $-4.24801$ & 6.6370 & W+P+G & $8.75^{+0.21}_{-0.27}$ & $4.1^{+1.8}_{-1.1}$ & $42.39^{+0.08}_{-0.10}$ & $41.08^{+0.16}_{-0.14}$ & $39.53^{+0.16}_{-0.14}$ & $44.39^{+0.63}_{-0.64}$ & $43.31^{+0.31}_{-0.32}$ \\
O3-09\tablenotemark{b} & 137.72726 & $-4.23208$ & 6.6270 & W+P+G & $9.59^{+0.14}_{-0.15}$ & $25.2^{+11.5}_{-9.1}$ & $42.35^{+0.08}_{-0.11}$ & $41.87^{+0.15}_{-0.18}$ & $40.32^{+0.16}_{-0.19}$ & $44.37^{+0.63}_{-0.64}$ & $43.60^{+0.30}_{-0.32}$ \\
O3-10 & 137.72810 & $-4.23555$ & 6.6560 & W+P+G & $9.03^{+0.18}_{-0.20}$ & $6.2^{+2.8}_{-1.6}$ & $42.39^{+0.08}_{-0.10}$ & $41.28^{+0.15}_{-0.13}$ & $39.71^{+0.16}_{-0.13}$ & $44.40^{+0.63}_{-0.64}$ & $43.26^{+0.31}_{-0.31}$ \\
O3-11 & 137.73102 & $-4.23663$ & 6.6310 & W+P+G & $9.18^{+0.19}_{-0.22}$ & $7.9^{+3.6}_{-2.5}$ & $42.28^{+0.09}_{-0.11}$ & $41.38^{+0.15}_{-0.16}$ & $39.82^{+0.16}_{-0.16}$ & $44.30^{+0.63}_{-0.64}$ & $43.12^{+0.31}_{-0.31}$ \\
O3-12 & 137.73575 & $-4.22681$ & 6.6400 & W+P & $8.49^{+0.19}_{-0.22}$ & $2.0^{+0.8}_{-0.4}$ & $41.93^{+0.10}_{-0.12}$ & $40.77^{+0.14}_{-0.10}$ & $39.22^{+0.14}_{-0.11}$ & $44.01^{+0.64}_{-0.64}$ & $43.15^{+0.31}_{-0.31}$ \\
O3-13 & 137.73894 & $-4.24618$ & 6.6330 & W+P+G & $9.00^{+0.17}_{-0.17}$ & $5.7^{+2.9}_{-1.5}$ & $42.39^{+0.08}_{-0.10}$ & $41.23^{+0.17}_{-0.12}$ & $39.68^{+0.18}_{-0.14}$ & $44.39^{+0.63}_{-0.64}$ & $43.37^{+0.31}_{-0.32}$ \\
O3-14 & 137.74239 & $-4.22685$ & 6.5650 & W & $9.91^{+0.16}_{-0.15}$ & $28.6^{+23.3}_{-13.4}$ & $42.24^{+0.09}_{-0.11}$ & $41.95^{+0.23}_{-0.25}$ & $40.37^{+0.26}_{-0.27}$ & $44.27^{+0.63}_{-0.64}$ & $43.24^{+0.31}_{-0.31}$ \\
O3-15 & 137.74831 & $-4.22918$ & 6.6360 & W+P+G & $9.09^{+0.19}_{-0.19}$ & $9.7^{+3.7}_{-2.1}$ & $42.55^{+0.08}_{-0.10}$ & $41.46^{+0.13}_{-0.11}$ & $39.91^{+0.14}_{-0.11}$ & $44.54^{+0.63}_{-0.64}$ & $43.58^{+0.31}_{-0.31}$ \\
O3-16\tablenotemark{a} & 137.75539 & $-4.25627$ & 6.6225 & G & $9.54^{+0.77}_{-0.56}$ & $12.2^{+37.9}_{-7.8}$ & $41.73^{+0.12}_{-0.16}$ & $41.62^{+0.63}_{-0.47}$ & $40.00^{+0.61}_{-0.45}$ & $43.84^{+0.64}_{-0.64}$ & $43.42^{+0.31}_{-0.32}$ \\
\enddata
\tablenotetext{a}{This source is also a spectroscopically confirmed LAE.}
\tablenotetext{b}{This source is also identified as an SMG.}
\tablecomments{W, P, and G denote usable JWST/NIRCam WFSS, JWST/NIRSpec PRISM, and JWST/NIRSpec G395M spectra, respectively. The listed mass and SFR are delayed-$\tau$ posterior medians with 16th--84th percentile statistical intervals; SFH systematics are discussed in the text and shown in Figure~\ref{fig:host_oiii}. $L_{5007}$ is observed and not corrected for dust attenuation within the host galaxy; its interval combines statistical and adopted systematic line-flux errors. The XRB and hot-gas intervals propagate the delayed-$\tau$ SED posterior through the rest-frame 2--10~keV relation of \citet{Lehmer2016ApJ} and the rest-frame 0.5--2~keV relation of \citet{Mineo2012MNRAS.426.1870M}, respectively; they do not include additional calibration scatter. {No metallicity-dependent correction is applied, and the listed XRB luminosities should therefore be regarded as nominal estimates.} The [O III]-based interval combines the line-measurement error with the 0.63~dex intrinsic dispersion of the AGN relation \citep{Berney2015MNRAS}, and the $\alpha_{\rm ox}$ interval combines the photometric error with an intrinsic $\alpha_{\rm ox}$ scatter of 0.12 \citep{Nanni2017A\string&A}. The final two columns are AGN-only limiting cases.}
\end{deluxetable*}

Both tests disfavor the hypothesis that unobscured, X-ray-normal AGN dominate the rest-frame UV continua and [\iona{O}{iii}] emission of all 16 galaxies. However, they do not exclude AGN contributions in individual galaxies or strongly obscured or intrinsically X-ray-weak AGN.

\subsection{Incidence of X-ray AGN activity in the \texorpdfstring{J0910$-$0414}{J0910-0414} protocluster}\label{sec:incidence_AGN}
After excluding the central quasar to avoid the selection bias inherent to a quasar-centered field, we detect no X-ray AGN among the 21 redshift-selected protocluster members ($6.53<z<6.73$; Section~\ref{sec:protocluster_individual}). For zero detections among 21 galaxies, the upper boundary of the central 90\% Jeffreys credible interval is 0.086 \citep{Brown01}; we therefore report $f_{\rm AGN}<0.09$.

\subsubsection{Comparison with lower-redshift protoclusters}
\begin{figure*}[htbp]
  \centering
  \includegraphics[width=\textwidth]{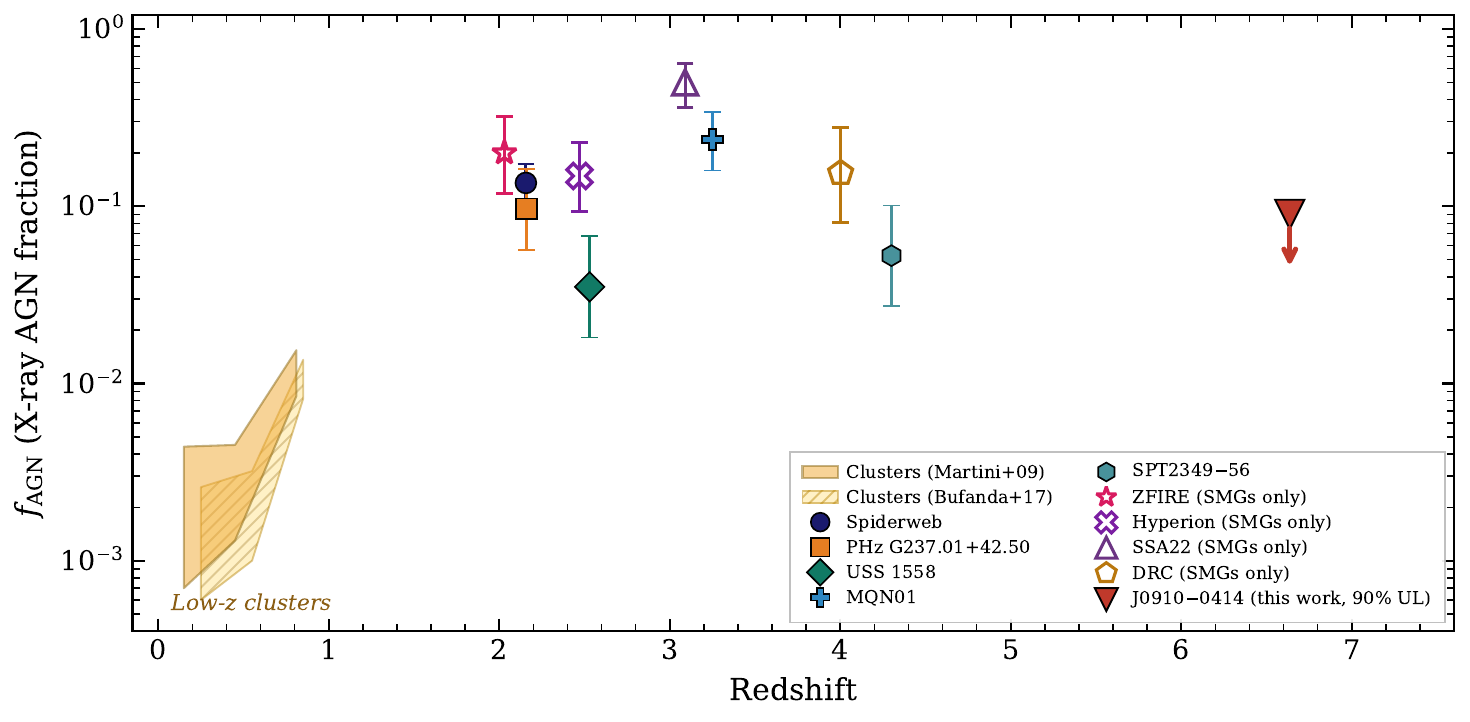}
  \caption{X-ray AGN fraction as a function of redshift for protoclusters with dedicated \chandra~observations. Each protocluster is plotted with a distinct marker symbol (see legend). Open symbols denote the published SMG/DSFG-only fractions for ZFIRE and Hyperion \citep{IslaLlave2026arXiv}, SSA22 \citep{Vito2024A&A,Umehata2019Sci,Lehmer2009ApJ}, and DRC \citep{Vito2020A&A}. All other protocluster fractions use the broader member samples defined in the cited studies. The red downward triangle marks the upper boundary of the central 90\% Jeffreys credible interval for J0910$-$0414 at $z=6.63$; error bars for the detected fractions show central 68\% Jeffreys credible intervals. The shaded bands at low redshift show the cluster X-ray AGN fractions from \citet{Martini2009ApJ} and \citet{Bufanda2017MNRAS}. The plotted points show the published fractions before applying a common X-ray luminosity threshold; the partial fixed-threshold comparison is described in the text. Because the member selections and X-ray sensitivities differ among structures, the figure provides descriptive context rather than a matched quantitative comparison. \label{fig:agnfrac}}
 \end{figure*}

Table~\ref{tbl:agnfrac} compiles X-ray AGN fractions for protoclusters at $z\approx2$--4 with dedicated \chandra\ observations, and Figure~\ref{fig:agnfrac} shows their redshift distribution. The denominators follow those reported in the cited studies. The published ZFIRE and Hyperion measurements are fractions among spectroscopically confirmed DSFGs \citep{IslaLlave2026arXiv}; the SSA22 value is the published SMG-only fraction \citep{Vito2024A&A}, based on the SMG sample of \citet{Umehata2019Sci} and the X-ray catalog of \citet{Lehmer2009ApJ}; and the DRC value is likewise an SMG-only fraction \citep{Vito2020A&A}. We retain these published denominators and show the four measurements with open symbols in Figure~\ref{fig:agnfrac}. All other structures use the broader member samples defined in their respective studies. The published fractions span approximately 4\%--50\%. On the basis of these values, the J0910$-$0414 upper limit lies near the low end of the distribution, but the samples differ substantially in X-ray depth and member selection and therefore cannot be compared directly.

We perform a partial fixed-threshold comparison to assess the effect of the heterogeneous X-ray sensitivities. We adopt $\log(L_{\rm X}/{\rm erg~s^{-1}})=43.8$, which matches the median limiting luminosity derived from the pixel-level sensitivity map within the adopted $8\arcmin$-radius aperture. At this threshold, we define the AGN fraction as the number of X-ray AGN brighter than the adopted limit divided by the number of members whose local X-ray limits are sufficiently deep to detect such an AGN. Using the individual 90\% luminosity upper limits as proxies for the local sensitivity, 18 of the 21 J0910$-$0414 members satisfy this requirement. The remaining three are LAE-only members located at relatively large off-axis angles, where the broader \chandra\ PSF results in shallower luminosity limits. Zero detections among the 18 sufficiently constrained galaxies changes the upper bound from $f_{\rm AGN}<0.09$ for 0/21 to $f_{\rm AGN}<0.10$ for 0/18, using the upper boundary of the central 90\% Jeffreys credible interval. For the literature samples, we use published source luminosities where available and luminosities inferred from catalog fluxes otherwise. Applying the same threshold changes the nominal fractions from 0.14 to 0.04 in Spiderweb (13/96 to 4/96), from 0.10 to 0.03 in PHz~G237 (3/31 to 1/31), from 0.04 to 0.02 in USS~1558 (2/57 to 1/57), from 0.50 to 0.25 in SSA22 (6/12 to 3/12), and from 0.24 to 0.14 in MQN01 (5/21 to 3/21). An equivalent reclassification is not possible for ZFIRE and Hyperion because individual X-ray luminosities are not tabulated \citep{IslaLlave2026arXiv}; DRC and SPT2349$-$56 are excluded because their representative luminosity limits are above the adopted threshold. Thus, X-ray depth lowers all five reclassified literature fractions, but by different amounts, and a substantial range of 0.02--0.25 remains. The data therefore provide no evidence for enhanced X-ray-detectable AGN activity in J0910$-$0414.

Member selection remains an additional limitation. SMG-selected samples often yield higher X-ray AGN fractions than samples of LAEs or H$\alpha$ emitters because X-ray AGN preferentially reside in massive galaxies \citep[e.g.,][]{Kauffmann2003MNRAS,Alexander2005ApJ,Digby2010MNRAS,Lehmer2009ApJ,Lehmer2013ApJ,Wang2013ApJ}. As shown in Section~\ref{sec:host_expected_xray}, the J0910$-$0414 denominator is dominated by modest-mass [\iona{O}{iii}] emitters and LAEs. Our upper limit therefore applies specifically to this spectroscopically confirmed, predominantly emission-line-selected sample and should not be generalized to the full galaxy population of the structure, particularly dusty galaxies absent from the current member census.
 
\begin{deluxetable}{lcccccc}
\tablecaption{X-ray AGN fractions in J0910$-$0414 and comparison protoclusters\label{tbl:agnfrac}}
\tablewidth{0pt}
\tablehead{
\colhead{Protocluster} &
\colhead{$z$} &
\colhead{$N_{\rm sample}$} &
\colhead{$N_{\rm AGN}$} &
\colhead{$f_{\rm AGN}$} &
\colhead{$L_{\rm X,lim}$}&
\colhead{Member sample}\\[-5pt]
\colhead{} & \colhead{} & \colhead{} & \colhead{} & \colhead{} & \colhead{($10^{42}~\mathrm{erg\,s^{-1}}$)} & \colhead{} \\[-5pt]
\colhead{(1)} & \colhead{(2)} & \colhead{(3)} & \colhead{(4)} & \colhead{(5)}&\colhead{(6)}&\colhead{(7)}
}
\startdata
ZFIRE$^{a}$             & 2.03 & 15 & 3  & $0.20^{+0.12}_{-0.08}$& $10$& SMG only\\
Spiderweb$^{b}$          & 2.16 & 96 & 13 & $0.14^{+0.04}_{-0.03}$& 6.8& LAE, HAE, SMG, CO \\
PHz G237$^{c}$          & 2.16 & 31 & 3  & $0.10^{+0.07}_{-0.04}$&11.0&HAE, SMG \\
Hyperion$^{a}$          & 2.47 & 27 & 4  & $0.15^{+0.08}_{-0.06}$& $15$& SMG only\\
USS 1558$^{d}$       & 2.53 & 57 & 2  & $0.04^{+0.03}_{-0.02}$&20.0& HAE (incl. SMGs) \\
SSA22$^{e}$              & 3.09 & 12 & 6  & $0.50^{+0.14}_{-0.14}$&5.2& SMG only\\
MQN01$^{f}$              & 3.25 & 21 & 5  & $0.24^{+0.10}_{-0.08}$&20.2& LAE, SMG, CO \\
DRC$^{g}$                & 4.00 & 13 & 2  & $0.15^{+0.12}_{-0.07}$&158.7& SMG only\\
SPT2349$-$56$^{h}$  & 4.30 & 38 & 2  & $0.05^{+0.05}_{-0.03}$ &101.0& LAE, SMG\\
\hline
J0910$-$0414$^{i}$       & 6.63 & 21 & 0  & $<0.09$ & 19.3& LAE, SMG, [O III], LBG\\
\enddata
\tablecomments{
Columns: (1) Protocluster name; (2) redshift; (3) number of objects in the adopted denominator; (4) number of X-ray-detected AGN; (5) X-ray AGN fraction; (6) representative rest-frame 2--10~keV luminosity limit; (7) member sample. Fractions with detections are reported with central 68\% Jeffreys credible intervals, whereas the J0910$-$0414 value is the upper boundary of the central 90\% Jeffreys credible interval. CO denotes carbon monoxide line emitters. For ZFIRE, Hyperion, SSA22, and DRC, $N_{\rm sample}$ includes only spectroscopically confirmed SMGs/DSFGs; for the other structures, it represents the broader member sample defined in the cited study. The approximate luminosity limits for ZFIRE and Hyperion are inferred from the survey-wide \chandra\ COSMOS-Legacy soft/full-band flux limits, assuming $\Gamma=1.4$ and the adopted cosmology, rather than reported directly by \citet{IslaLlave2026arXiv}. For J0910$-$0414, the value in Column~(6) is the minimum 2.6$\sigma$ field flux limit and does not represent a uniform limit for all 21 members; their individual 90\% luminosity upper limits span $(21.6$--$133)\times10^{42}~\mathrm{erg\,s^{-1}}$ and are used in the partial fixed-threshold test described in the text. \\
References: $^{a}$\citet{IslaLlave2026arXiv}, with approximate luminosity limits inferred from \citet{Civano2016ApJ}; $^{b}$\citet{Tozzi2022A\string&A}; $^{c}$\citet{Polletta2021A\string&A}; $^{d}$\citet{Macuga2019ApJ}; $^{e}$\citet{Vito2024A\string&A}, based on the SMG sample of \citet{Umehata2019Sci} and X-ray catalog of \citet{Lehmer2009ApJ}; $^{f}$\citet{Travascio2025AA}; $^{g}$\citet{Vito2020A\string&A}; $^{h}$\citet{Vito2024A\string&A}; $^{i}$This work.
}
\end{deluxetable}

\subsubsection{Search for additional high-redshift X-ray sources}
The incidence measurement above is limited to the 21 redshift-selected members with secure spectroscopic redshifts. To test whether an X-ray-bright member could have been excluded from that spectroscopic census, we also search the full catalog of 251 X-ray sources for additional candidates at $z\sim6.6$.

Following standard $z\sim6$ $i$-dropout selections \citep[e.g.,][]{DowHygelund2007ApJ}, and adapting them to the available optical catalog, we require non-detections in $g$, $r$, and $i$ ($<2\sigma$), a detection in $z$ ($>5\sigma$), and $(i-z)>1.3$. Only XID~29, the central quasar, satisfies these criteria. We separately examined the sources lacking HSC and DESI Legacy optical counterparts. XIDs~71, 73, 109, and 249 are detected with JWST and/or HST, while XIDs~125 and 213 have near-infrared VHS counterparts. The available multiband imaging, WFSS grism spectra where available, photometric-redshift fits with EAZY \citep{EAZY2008ApJ} and CIGALE \citep{CIGALE2019A&A}, and broadband SED inspection provide no robust evidence that any of these sources lies at $z>6$.

For the four catalog sources that remain without identified multiwavelength counterparts, two lie behind bright foreground stars, preventing host identification, and two fall outside the JWST/HST imaging footprint. Their redshifts remain unconstrained, so association with the protocluster cannot be formally excluded.

Within the $8\arcmin$ aperture and the interval $6.53<z<6.73$, the field XLF predicts only $N_{\rm exp,field}=0.0356$ X-ray AGN (Section~\ref{sec:space_density}), making a chance field source in this narrow redshift slice unlikely. Even after scaling by the measured galaxy overdensity, the corresponding expectation is only $N_{\rm exp,OD}=0.131$ and remains small. For comparison, the 7~Ms CDF-S catalog likewise contains no X-ray source with an adopted redshift $z>6$ \citep{Luo2017ApJS..228....2L}. We conclude that no cataloged X-ray source other than the central quasar can be robustly associated with the J0910$-$0414 protocluster, while the four sources without usable host constraints remain unconstrained.

\subsection{Field X-ray AGN expectation and overdensity-corrected enhancement constraint}
\label{sec:space_density}
We compare the absence of detected companion X-ray AGN in J0910$-$0414 with the field expectation following the approach of \citet{Vito2024A&A}. The calculation uses the $8\arcmin$-radius aperture and the adopted protocluster interval $6.53<z<6.73$, which correspond at $z=6.63$ to a comoving radius of $19.8~\mathrm{cMpc}$, a line-of-sight depth of $74.0~\mathrm{cMpc}$, and a survey volume of $9.09\times10^4~\mathrm{cMpc}^3$.

For the field population, we adopt the redshift-dependent XLF model of \citet{Pouliasis2024A&A} for intrinsic, absorption-corrected rest-frame 2--10~keV emission from Compton-thin AGN. The model is calibrated over $3\leq z\leq6$ and is extrapolated to $z=6.63$ here. We account for the nonuniform \chandra\ depth by weighting the XLF with the sky coverage derived from the pixel-level full-band sensitivity map. The flux-to-luminosity conversion assumes an unabsorbed power law with $\Gamma=1.8$, as adopted for the member-source limits in Section~\ref{sec:protocluster_individual}. The resulting 16th, 50th, and 84th percentiles of the limiting luminosity at $z=6.63$ are $\log(L_{\rm X}/{\rm erg~s^{-1}})=43.58$, 43.79, and 44.01, respectively.

The sensitivity-weighted calculation predicts $N_{\mathrm{exp,field}}=0.0356$ field X-ray AGN in the surveyed volume. {Accounting for the uncertainties in the XLF parameters gives a range of $N_{\mathrm{exp,field}}=0.0245$--$0.0500$.} With the central quasar excluded, zero detected companions corresponds to a one-sided 90\% Poisson upper limit of $N_{\mathrm{AGN}}<2.303$ \citep{Gehrels1986ApJ...303..336G}. We first define the raw density-enhancement factor as the ratio of the protocluster to field X-ray AGN space density, $\eta_{\rm dens}=N_{\mathrm{AGN}}/N_{\mathrm{exp,field}}$. The non-detection gives $\eta_{\rm dens}<64.7$ for the best-fit XLF, {with a corresponding range of $\eta_{\rm dens}<46$--94.} This comparison does not account for the enhanced number density of potential AGN hosts in the protocluster.

\begin{figure*}[htbp]
\centering
\includegraphics[width=\textwidth]{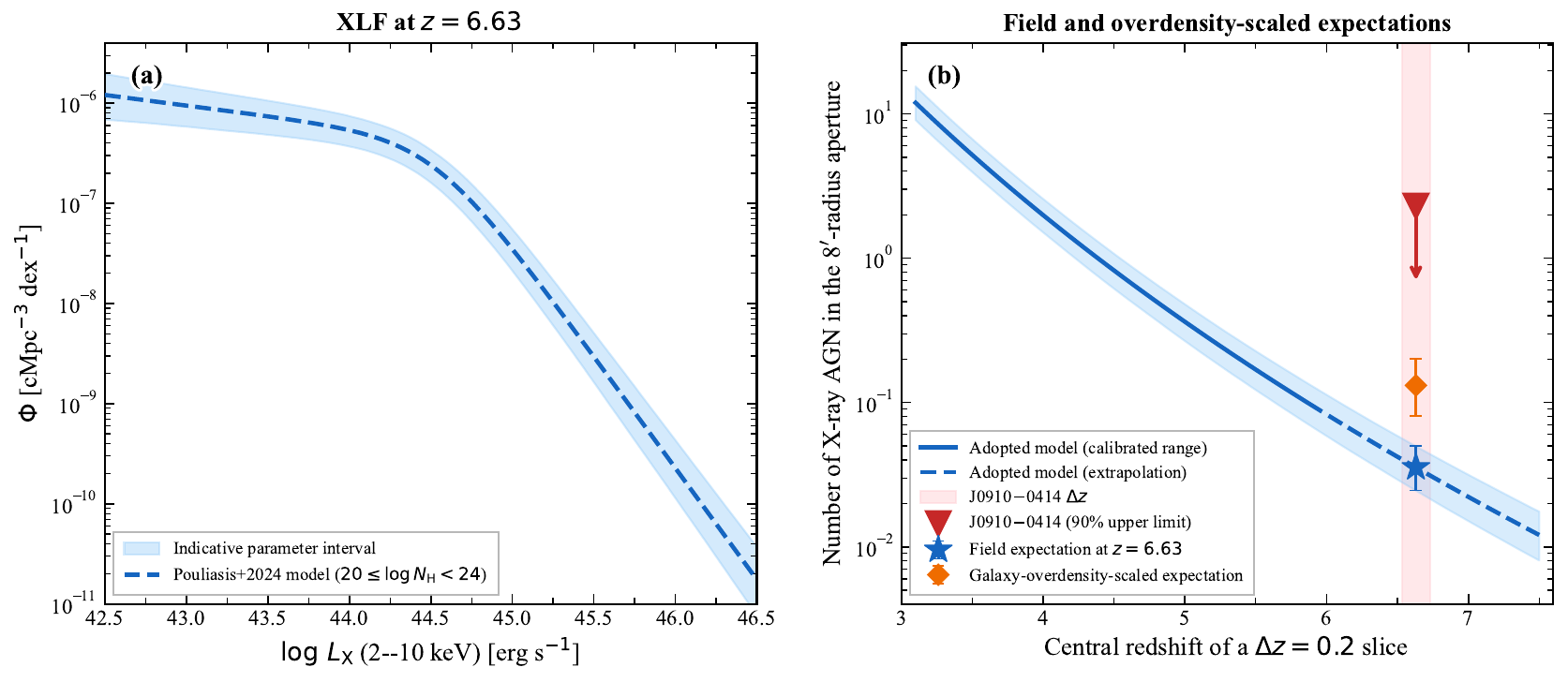}
\caption{{(a)} Compton-thin XLF at $z=6.63$ from the adopted model of \citet{Pouliasis2024A&A}. {The blue band includes the uncertainties in the XLF parameters.} {(b)} Expected number of X-ray AGN in the $8\arcmin$-radius aperture and a redshift slice of width $\Delta z=0.2$, after accounting for the spatially varying 2.6$\sigma$ sensitivity. Solid and dashed curves distinguish the calibrated ($3\leq z\leq6$) and extrapolated ($z>6$) field-XLF regimes. The blue star marks the field expectation, $N_{\rm exp,field}=0.0356$, at $z=6.63$, {and its error bar includes the same uncertainties.} The orange diamond marks the overdensity-scaled expectation, $N_{\rm exp,OD}=0.131$, and its error bar additionally includes the uncertainty in the galaxy overdensity. The red downward triangle marks the 90\% upper limit from zero detected companions, and the pink shaded region indicates the adopted protocluster redshift interval.
\label{fig:xlf_space_density}}
\end{figure*}

To isolate any X-ray AGN enhancement beyond the galaxy overdensity, we use the cumulative LAE density profile measured by \citet{Wang2024ApJ...962L..11W}. At $R=20.48~\mathrm{cMpc}$, which closely matches the projected radius of the $8\arcmin$ aperture, the profile gives $1+\delta=3.69^{+1.17}_{-0.91}$. Assuming that this density ratio is representative of the full adopted redshift interval, the expected count for a field-like X-ray AGN incidence per galaxy is $N_{\rm exp,OD}=(1+\delta)N_{\rm exp,field}=0.131$. {Accounting for the uncertainties in the XLF parameters and the galaxy overdensity gives a range of $N_{\rm exp,OD}=0.081$--$0.201$.} We define the additional enhancement relative to this overdensity-scaled expectation as $\eta_{\rm rel}=N_{\rm AGN}/N_{\rm exp,OD}=\eta_{\rm dens}/(1+\delta_{\rm gal})$. The non-detection gives $\eta_{\rm rel}<17.5$, {with a corresponding range of $\eta_{\rm rel}<11.5$--28.5.} We therefore find no evidence for an additional enhancement of X-ray-detectable AGN beyond that associated with the galaxy overdensity, although the limit still permits more than an order-of-magnitude enhancement and is not yet stringent. Figure~\ref{fig:xlf_space_density} shows both the field and overdensity-scaled expectations.

Because the adopted XLF describes Compton-thin AGN and is extrapolated slightly beyond its calibrated redshift range, the resulting limits apply only to the X-ray-detectable Compton-thin population, not to the total incidence of SMBH accretion. The cumulative LAE profile was measured over $|\Delta v|<2250~\mathrm{km~s^{-1}}$, which is narrower than the full $\Delta z=0.2$ interval adopted here. The overdensity correction also assumes that LAEs trace the relevant population of potential X-ray AGN hosts. Therefore, $\eta_{\rm rel}$ should be regarded as an approximate environment-normalized constraint under these assumptions.

\subsection{Interpretation of the X-ray non-detection}
\label{sec:nondetection_discussion}
The absence of detected X-ray AGN among the J0910$-$0414 members does not by itself imply an absence of SMBH accretion. Heavy obscuration, unusually soft intrinsic spectra, intermittent activity, and intrinsic X-ray weakness can all reduce the probability that an accreting SMBH satisfies our X-ray selection.

Obscuration is likely to be important at these redshifts. X-ray population studies at $3\leq z\leq6$ infer that $61^{+8}_{-7}\%$ of AGN have $\log(N_{\rm H}/{\rm cm^{-2}})\geq23$, while the inferred Compton-thick fraction is $f_{\rm CTK}=17^{+7}_{-9}\%$ \citep{Vito2018MNRAS.473.2378V,Pouliasis2024A&A}. For comparison, the warm-infrared-selected NuLANDS sample of nearby Seyfert galaxies yields a higher Compton-thick fraction of $35\%\pm9\%$ at 90\% confidence \citep{Boorman2025ApJ}; the different selection functions, luminosity distributions, and spectral models preclude interpreting this difference as redshift evolution. In the $\gamma=2$ host-ISM model of \citet{Gilli2022A&A}, more than 80\% of SMBHs at $z>6$ are obscured by their host galaxies, while Compton-thick ISM absorption affects approximately 15--20\% of accreting SMBHs at $z>8$. Cosmic X-ray background synthesis models likewise predict a substantial Compton-thick population that is missed by current X-ray surveys \citep{Gilli2007A&A,Treister2009ApJ}.

To quantify the effect on our observations, we simulated the 0.5--7~keV attenuation at $z=6.63$ with Sherpa \citep{Siemiginowska2024arXiv}, adopting a $\Gamma=1.8$ power law and the model \textsc{phabs}$\times$\textsc{cabs}$\times$\textsc{zphabs}$\times$\textsc{powerlaw}. The observed band corresponds to rest-frame $\sim3.8$--53~keV, which reduces photoelectric absorption but not Compton-scattering losses. The observed flux is suppressed by $\approx72\%$ at $N_{\rm H}=10^{24}~\mathrm{cm^{-2}}$, by $\approx99\%$ at $5\times10^{24}~\mathrm{cm^{-2}}$, and is effectively extinguished at $10^{25}~\mathrm{cm^{-2}}$. At $N_{\rm H}=10^{24}~\mathrm{cm^{-2}}$, the sample-averaged stacking limit of $L_{\rm 2-10~keV}\leq9.54\times10^{42}~\mathrm{erg~s^{-1}}$ corresponds to an intrinsic limit of $L_{\rm 2-10~keV}\lesssim3.4\times10^{43}~\mathrm{erg~s^{-1}}$. A population in which every member hosted an unabsorbed AGN with $L_{\rm 2-10~keV}\sim10^{43}~\mathrm{erg~s^{-1}}$ would approach the mean stacking limit. Because the stack averages over 21 galaxies, however, moderate-luminosity AGN hosted by only a subset of the members could remain undetected even without attenuation, and obscuration would weaken the constraint further.

An unusually soft intrinsic spectrum would further reduce detectability. At $z=6.63$, rest-frame 2--10~keV corresponds to observed-frame $\approx0.26$--1.31~keV, so a larger fraction of a steep spectrum falls below the 0.5~keV lower boundary of our analysis. \citet{Connor2020ApJ}, for example, measured $\Gamma=2.6^{+1.0}_{-0.9}$ for a quasar at $z=6.59$. Limits derived assuming $\Gamma=1.8$ do not capture this less favorable spectral shape.

The measurement also constrains only the instantaneous fraction of members above the adopted X-ray luminosity threshold. Short or intermittent accretion episodes may therefore remain undetected even if they contribute substantially to cumulative SMBH growth \citep{Hickox2014ApJ,Blecha2018MNRAS}. In addition, JWST has revealed moderate-luminosity AGN candidates, including little red dots, that are undetected or substantially weaker in X-rays than expected from conventional AGN scaling relations \citep{Matthee2024ApJ,Ananna2024ApJL,Yue2024ApJL,Maiolino2025MNRAS}. Such sources may be underrepresented in X-ray-derived luminosity functions because of heavy obscuration and/or intrinsic X-ray weakness. The current data cannot distinguish a genuinely low total AGN incidence from a substantial X-ray-faint population. They constrain only the incidence of X-ray-detectable AGN above our luminosity threshold.

\section{Summary and conclusions}\label{sec:5}
We have presented deep \chandra\ observations ($\sim865$~ks) of the J0910$-$0414 field, which contains a massive spectroscopically confirmed protocluster at $z=6.63$ \citep{Wang2024ApJ...962L..11W} anchored by a luminous broad absorption-line (BAL) quasar \citep{Yang2021ApJ}.  Our main results are as follows:

\begin{enumerate}[noitemsep, topsep=3pt]

\item We constructed a catalog of 251 sources detected in at least one of the full (0.5--7~keV), soft (0.5--2~keV), and hard (2--7~keV) bands; 247 are detected in the full band. The minimum 2.6$\sigma$ full-band flux limit is $1.34\times10^{-16}~\mathrm{erg~s^{-1}~cm^{-2}}$. We identified secure multiwavelength counterparts for 227 sources (90.4\%) and plausible counterparts for a further 20 (8.0\%). The cumulative number counts are broadly consistent with the empirical broken-power-law relation of \citet{Georgakakis2008MNRAS.388.1205G} and measurements from other deep fields. See Section~\ref{sec:2}.

\item None of the 21 spectroscopically confirmed galaxies in the adopted protocluster redshift interval is individually detected in X-rays; their 90\% upper limits span $L_{\rm 2-10~keV}\lesssim(2$--$13)\times10^{43}~\mathrm{erg~s^{-1}}$. Stacking all 21 galaxies ($\sim17$~Ms total exposure) yields no significant signal and constrains the average luminosity to $L_{\rm 2-10~keV}\leq9.54\times10^{42}~\mathrm{erg~s^{-1}}$. See Sections~\ref{sec:protocluster_individual} and \ref{sec:stack}.

\item For the 16 [\iona{O}{iii}] members, joint fits to the total broad-band photometry and independently measured [\iona{O}{iii}] $\lambda5007$ fluxes give a median stellar mass of $\log(M_\star/M_\odot)=9.05$ and a median SFR of $7.1~M_\odot\,\mathrm{yr}^{-1}$, with substantial SFH-dependent systematics. The exposure-weighted 2--10~keV XRB prediction is $2.77\times10^{41}~\mathrm{erg\,s^{-1}}$, about 16 times below the corresponding stacking limit, while the predicted diffuse 0.5--2~keV luminosity is only $7.6\times10^{39}~\mathrm{erg\,s^{-1}}$; the non-detections therefore do not constrain either normal-galaxy component. We also consider two AGN-only limiting cases. If all [\iona{O}{iii}] emission is attributed to AGN activity, the predicted luminosities exceed every individual X-ray limit. If the total UV continuum is instead attributed to an unobscured type-1 AGN, the exposure-weighted mean predicted luminosity is 4.45 times higher than the stacking limit. These limiting cases disfavor domination of the full subsample by unobscured, X-ray-normal AGN but do not classify individual galaxies. 

\item The upper boundary of the central 90\% Jeffreys credible interval on the X-ray AGN fraction is $f_{\rm AGN}<0.09$. In the partial fixed-threshold comparison at $\log(L_{\rm X}/\mathrm{erg~s^{-1}})=43.8$, the limit becomes $f_{\rm AGN}<0.10$ for zero detections among 18 sufficiently sensitive members, while the reclassified literature fractions span 0.02--0.25. The data provide no evidence for an enhanced X-ray AGN fraction in J0910$-$0414. See Section~\ref{sec:incidence_AGN}.

\item Folding the field X-ray luminosity function through the survey sensitivity \citep{Pouliasis2024A&A}, we find $N_{\rm exp,field}=0.0356$ expected field X-ray AGN in the $8\arcmin$ aperture and $\Delta z=0.2$ slice. Scaling by the observed galaxy overdensity in this protocluster at this scale, $1+\delta=3.69^{+1.17}_{-0.91}$, we derive $N_{\rm exp,OD}=0.131$ under the assumption of AGNs have the same overdensity profile as galaxies. The non-detection constrains any additional enhancement beyond the galaxy overdensity to $\eta<17.5$, {with a range of $\eta<11.5$--28.5 after accounting for the uncertainties in the XLF parameters and the galaxy overdensity.} We therefore find no evidence for an additional enhancement of X-ray-detectable AGN activity, although the constraint still permits more than an order-of-magnitude enhancement. See Section~\ref{sec:space_density}.

\item The non-detection does not distinguish a genuinely low AGN incidence from heavy obscuration, unusually steep X-ray spectra that shift emission toward or below the low-energy edge of the \chandra\ bandpass, intrinsic X-ray weakness, or short-lived and intermittent X-ray-bright accretion.  See Section~\ref{sec:nondetection_discussion}.

\end{enumerate}

To our knowledge, this work provides the first systematic X-ray census of AGN activity among member galaxies in a $z>6$ protocluster. After normalizing the field XLF expectation by the observed galaxy overdensity, we find no evidence for an additional enhancement of X-ray-detectable AGN and constrain it to $\eta_{\rm rel}<17.5$ at 90\% confidence. This upper limit is not sufficiently stringent to determine why no member AGN is detected. Deeper X-ray observations of additional spectroscopically confirmed $z>6$ protoclusters will be needed to distinguish obscured or intrinsically X-ray-weak AGN from short-lived X-ray-bright accretion phases or a genuinely low overall AGN incidence. Broader member censuses, including submillimeter-selected dusty galaxies, will also be required to determine whether constraints from predominantly emission-line-selected samples extend to the full protocluster galaxy population.

Next-generation X-ray observatories such as NewAthena will offer effective areas at least an order of magnitude larger than that of Chandra at 1 keV, substantially reducing the exposure times required to characterize samples of high-redshift protoclusters such as J0910$-$0414 \citep{Cruise2025NatAs}.

\begin{acknowledgments}
Support for this work was partially provided by the National Aeronautics and Space Administration through Chandra Award Number 23700397 issued by the Chandra X-ray Center, which is operated by the Smithsonian Astrophysical Observatory for and on behalf of the National Aeronautics and Space Administration under contract NAS8-03060. CM acknowledges support from Fondecyt Iniciacion grant 11240336 and the ANID BASAL project FB210003. FV acknowledges support from the “INAF Ricerca Fondamentale 2023 - Large GO” grant. J.T.L. acknowledges the financial support by the National Natural Science Foundation of China (NSFC) through grants 12321003  and 12273111, from the China Manned Space Program through grants CMS-CSST-2025-A10 and CMS-CSST-2025-A04, and from the Jiangsu Innovation and Entrepreneurship Talent Team Program through grant JSSCTD202436.

We thank Xiurui Zhao for helpful discussions.
The scientific results reported in this article are based on 
observations made by the Chandra X-ray Observatory. This research has made use of software provided 
by the Chandra X-ray Center (CXC) in the application packages 
CIAO (version 4.17) and Sherpa. {The Chandra observations analyzed in this work can be accessed via the Chandra Data Collection \dataset[doi:10.25574/cdc.660]{https://doi.org/10.25574/cdc.660}.}

This work is also based on observations made with the NASA/ESA Hubble Space Telescope and obtained from the Mikulski Archive for Space Telescopes at the Space Telescope Science Institute, which is operated by the Association of Universities for Research in Astronomy, Inc., under NASA contract NAS 5-26555. These observations are associated with program GO-16187 and can be accessed via \dataset[doi:10.17909/hq0m-fs75]{https://doi.org/10.17909/hq0m-fs75}.

This work is partially based on observations made with the NASA/ESA/CSA James Webb Space Telescope. The data were obtained from the Mikulski Archive for Space Telescopes at the Space Telescope Science Institute, which is operated by the Association of Universities for Research in Astronomy, Inc., under NASA contract NAS 5-03127 for JWST. These observations are associated with program \#2078 and \#2028 and can be accessed via \dataset[doi:10.17909/vt74-kd84]{https://doi.org/10.17909/vt74-kd84} and \dataset[doi:10.17909/1ccb-z628]{https://doi.org/10.17909/1ccb-z628}, respectively.

This paper makes use of the following ALMA data: 
ADS/JAO.ALMA\#2018.1.01188.S, 
ADS/JAO.ALMA\#2019.1.01216.S, and 
ADS/JAO.ALMA\#2022.1.01077.L. ALMA is a partnership of ESO (representing its member states), NSF (USA) and NINS (Japan), together with NRC (Canada), MOST and ASIAA (Taiwan), and KASI (Republic of Korea), in cooperation with the Republic of Chile. The Joint ALMA Observatory is operated by ESO, AUI/NRAO and NAOJ. The National Radio Astronomy Observatory is a facility of the National Science Foundation operated under cooperative agreement by Associated Universities, Inc.

This research used \textit{Gaia} DR3 products generated by the Gaia Data Processing and Analysis Consortium for the European Space Agency; the consortium is supported by the national institutions participating in the Gaia Multilateral Agreement. The \textit{GALEX} catalog was obtained from the Mikulski Archive for Space Telescopes (MAST). This work also used imaging from the DESI Legacy Imaging Surveys and unWISE products incorporating data from NASA's Wide-field Infrared Survey Explorer and NEOWISE missions.

The HSC data were collected with the Subaru Telescope, which is operated by the National Astronomical Observatory of Japan. We acknowledge the cultural significance of Maunakea to the Native Hawaiian community.

The VHS data are based on observations collected at the European Southern Observatory under program 179.A-2010 (PI: R. McMahon). The data were processed with the VISTA Data Flow System at the Cambridge Astronomy Survey Unit and accessed through the VISTA Science Archive \citep{Irwin2004SPIE.5493..411I,Lewis2010ASPC..434...91L,Cross2012A&A...548A.119C}.

\end{acknowledgments}
\software{
CIAO \citep{CIAO2006SPIE.6270E..1VF},
Sherpa \citep{Siemiginowska2024arXiv},
Astropy \citep{Astropy2022ApJ...935..167A},
NWAY \citep{Salvato2018MNRAS.473.4937S},
\textsc{bagpipes} \citep{Carnall2018MNRAS},
FastHR \citep{Zou2023ApJ...950..136Z},
FastSpecFit \citep{Moustakas2023ascl.soft08005M},
\textsc{eazy} \citep{EAZY2008ApJ},
\textsc{cigale} \citep{CIGALE2019A&A},
DrizzlePac \citep{DrizzlePac},
SExtractor \citep{Bertin1996A&AS},
SExtractor++ \citep{Bertin2020ASPC}
}

\appendix

\section{Detailed X-ray survey and catalog construction}
\label{app:catalog}

{This appendix provides the detailed procedures used to construct the X-ray source catalog summarized in Section~\ref{sec:2}.}

\subsection{Observation log, data reduction, and astrometry}

All observations used the ACIS imaging array in VFAINT telemetry mode. The ACIS-S2 CCD was active during 24 observations, but the analysis uses only events recorded on the four ACIS-I CCDs. The survey covers approximately $515~\mathrm{arcmin^2}$, and the exposure-weighted mean aim point is $(\mathrm{R.A.},\mathrm{Decl.})=(137.730^\circ,-4.239^\circ)$.

\begin{deluxetable*}{cccccccccccc}
\tablecaption{Summary of the Chandra observations of the \hbox{J0910$-$0414} field\label{tab:obslog}}
\tablewidth{0pt}
\tabletypesize{\scriptsize}
\setlength{\tabcolsep}{2pt}
\tablehead{
\colhead{OBSID} & \colhead{Start Date} & \colhead{$T_{\mathrm{exp}}$} & \colhead{R.A.} & \colhead{Decl.} & \colhead{Roll Angle} &
\colhead{OBSID} & \colhead{Start Date} & \colhead{$T_{\mathrm{exp}}$} & \colhead{R.A.} & \colhead{Decl.} & \colhead{Roll Angle} \\[-5pt]
\colhead{} & \colhead{} & \colhead{(ks)} & \colhead{(deg)} & \colhead{(deg)} & \colhead{(deg)} &
\colhead{} & \colhead{} & \colhead{(ks)} & \colhead{(deg)} & \colhead{(deg)} & \colhead{(deg)}
}
\startdata
25391 & 2022-11-17 & 28.70 & 137.73563 & $-4.22136$ & 65.21 & 25887 & 2023-03-19 & 29.69 & 137.72774 & $-4.25142$ & 276.32 \\
25865 & 2023-10-15 & 13.71 & 137.73109 & $-4.21981$ & 80.20 & 25888 & 2022-12-06 & 29.68 & 137.73587 & $-4.22137$ & 64.21 \\
25866 & 2023-02-04 & 11.57 & 137.74158 & $-4.23713$ & 1.01 & 25889 & 2023-05-09 & 29.67 & 137.72113 & $-4.24979$ & 253.23 \\
25867 & 2023-10-15 & 12.08 & 137.73036 & $-4.21918$ & 82.84 & 25890 & 2023-12-20 & 8.97 & 137.73669 & $-4.22219$ & 58.23 \\
25868 & 2023-03-14 & 15.65 & 137.73185 & $-4.25047$ & 292.19 & 25891 & 2023-01-27 & 16.81 & 137.74251 & $-4.23498$ & 8.20 \\
25869 & 2023-01-31 & 13.59 & 137.74103 & $-4.23822$ & 356.21 & 25892 & 2023-02-26 & 21.97 & 137.73438 & $-4.24958$ & 301.17 \\
25870 & 2023-05-22 & 19.83 & 137.72059 & $-4.25030$ & 251.20 & 25893 & 2022-06-25 & 24.75 & 137.71682 & $-4.24609$ & 235.00 \\
25871 & 2023-01-18 & 26.95 & 137.74005 & $-4.22862$ & 35.29 & 27676 & 2023-01-30 & 9.23 & 137.74141 & $-4.23242$ & 18.20 \\
25872 & 2023-03-12 & 31.67 & 137.72922 & $-4.25118$ & 282.20 & 27677 & 2023-01-28 & 12.74 & 137.74178 & $-4.23674$ & 2.21 \\
25873 & 2022-11-23 & 32.44 & 137.73447 & $-4.22055$ & 68.89 & 27683 & 2023-02-05 & 9.95 & 137.74102 & $-4.23444$ & 13.22 \\
25874 & 2023-05-14 & 18.83 & 137.71989 & $-4.24922$ & 249.20 & 27684 & 2023-02-05 & 9.95 & 137.74085 & $-4.23446$ & 13.21 \\
25875 & 2023-10-31 & 11.26 & 137.73006 & $-4.21931$ & 83.20 & 27685 & 2023-01-31 & 16.76 & 137.74166 & $-4.23442$ & 10.36 \\
25876 & 2023-05-06 & 31.47 & 137.72114 & $-4.25008$ & 254.04 & 27702 & 2023-02-11 & 14.72 & 137.74065 & $-4.24093$ & 345.20 \\
25877 & 2023-05-17 & 30.68 & 137.71954 & $-4.24966$ & 247.20 & 27713 & 2023-03-05 & 20.90 & 137.73198 & $-4.25088$ & 290.74 \\
25878 & 2023-05-03 & 28.21 & 137.72178 & $-4.25075$ & 255.17 & 27744 & 2023-04-27 & 19.15 & 137.72233 & $-4.25099$ & 257.19 \\
25879 & 2023-01-07 & 29.69 & 137.73969 & $-4.22446$ & 46.77 & 27845 & 2023-05-19 & 14.88 & 137.71906 & $-4.24875$ & 246.20 \\
25880 & 2023-02-06 & 17.88 & 137.74094 & $-4.23966$ & 350.21 & 27854 & 2023-05-22 & 14.91 & 137.72070 & $-4.25018$ & 251.20 \\
25881 & 2023-05-05 & 21.81 & 137.72136 & $-4.25051$ & 254.53 & 27855 & 2023-05-22 & 14.99 & 137.72067 & $-4.25050$ & 251.20 \\
25882 & 2023-01-29 & 17.18 & 137.74178 & $-4.23661$ & 2.21 & 29015 & 2023-10-31 & 11.43 & 137.73001 & $-4.21915$ & 83.20 \\
25883 & 2023-05-05 & 27.71 & 137.72142 & $-4.24994$ & 254.29 & 29016 & 2023-10-30 & 11.19 & 137.73020 & $-4.21941$ & 83.20 \\
25884 & 2023-10-19 & 9.65 & 137.73082 & $-4.21999$ & 81.25 & 29017 & 2023-10-30 & 11.36 & 137.73027 & $-4.21978$ & 83.19 \\
25885 & 2023-03-12 & 22.78 & 137.72938 & $-4.25155$ & 282.20 & 29018 & 2023-11-01 & 11.41 & 137.72998 & $-4.21900$ & 83.20 \\
25886 & 2023-01-12 & 27.03 & 137.73908 & $-4.22327$ & 50.21 & & & & & & \\
\enddata
\tablecomments{The \hbox{J0910$-$0414} field contains 45 observations, with a total cleaned exposure time of 865.5 ks. The six columns are repeated in the left and right blocks: Observation ID, start date, cleaned exposure time, Right Ascension, Declination, and spacecraft roll angle.}
\end{deluxetable*}
\onecolumngrid

We reprocessed the data with the \textit{chandra\_repro} script in CIAO 4.17 \citep{CIAO2006SPIE.6270E..1VF}, enabling \texttt{check\_vf\_pha=yes} for the VF mode data, and retained events with ASCA grades 0, 2, 3, 4, and 6. We then ran \textit{deflare} and applied iterative $3\sigma$ clipping. The observations were not significantly affected by background flares, and the total cleaned exposure is 865.5~ks.

We corrected the astrometry with \textit{fine\_astro}, following the CIAO thread. We first used \textit{merge\_obs} to reproject the observations to a common tangent point and generated an image for each observation. We ran \textit{wavdetect} with a significance threshold of $10^{-6}$ and retained point sources with a PSF size of $\leq3\arcsec$ and at least five counts. Sources brighter than $r=23.5$ in the DESI Legacy Imaging Survey DR10 catalog \citep{DESI2019AJ....157..168D} provided the astrometric reference list. We matched the X-ray sources to this list and determined translational offsets with \textit{wcs\_match}, then applied the offsets to the event and aspect solution files.

We ran \textit{wavdetect} on the merged full-band image and matched the resulting point sources to \textit{Gaia} DR3 \citep{Gaia2023A&A...674A...1G}. Eighteen unique one-to-one matches lie within $1\arcsec$. The median signed offsets, defined as X-ray minus \textit{Gaia}, are $\Delta\mathrm{R.A.}\cos(\mathrm{Decl.})=-0.03\arcsec$ and $\Delta\mathrm{Decl.}=+0.01\arcsec$; the median radial separation is $0.26\arcsec$, and 68\% of the matches lie within $0.44\arcsec$. A 1000-trial shifted-position test, using random displacements of 10--30\arcsec, predicts 0.46 chance associations within the adopted matching radius.

\begin{figure}[htbp]
\centering
\includegraphics[width=3.3in]{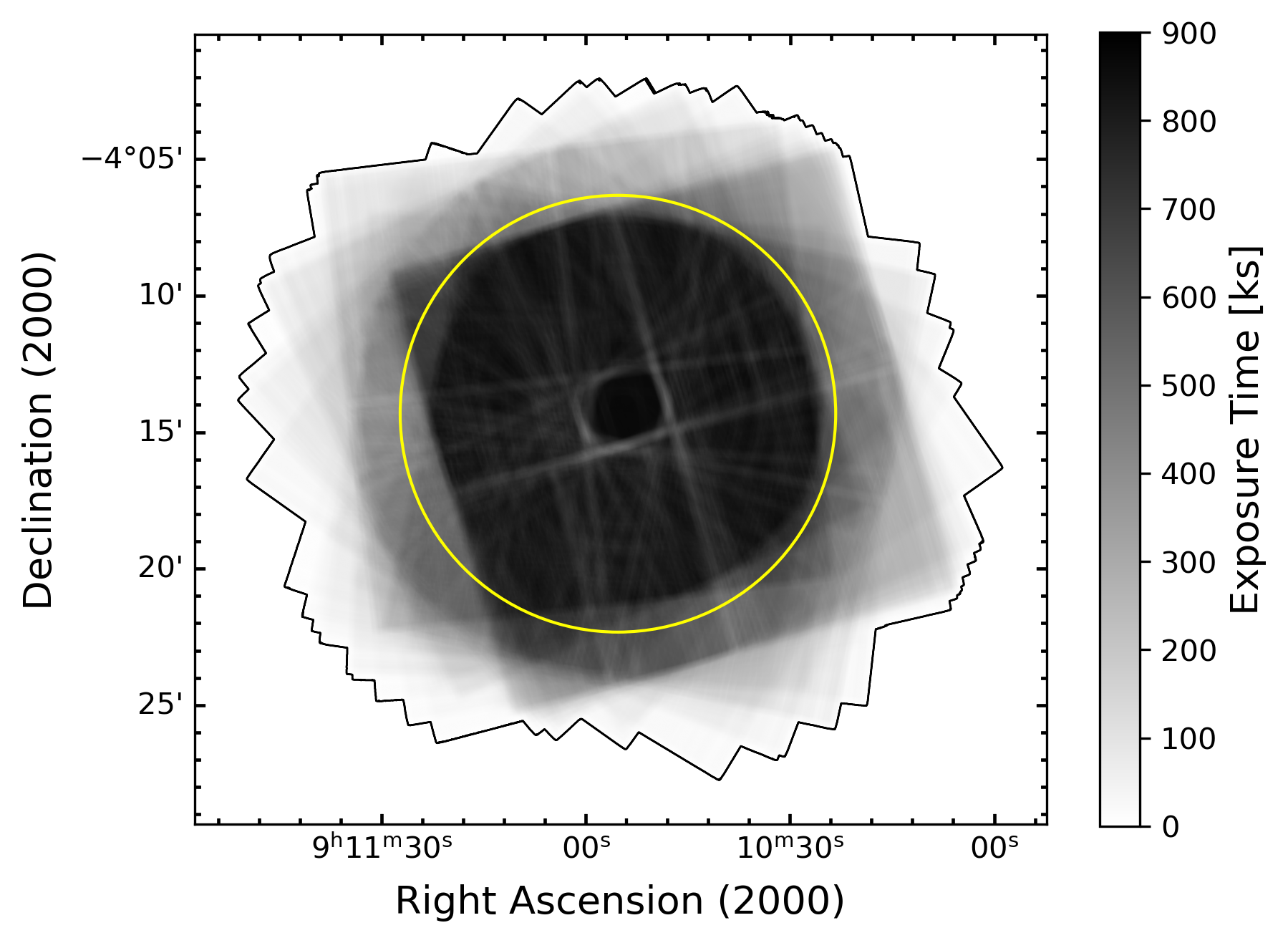}
\caption{Full-band exposure map of the J0910$-$0414 field. The linear gray scale bar shows the exposure time. The radial trails and central ring-like structure with relatively low effective exposures are caused by the ACIS-I CCD gaps. The yellow circle has a radius of $8\arcmin$ and marks the region adopted for the protocluster analysis. \label{fig:xexpmap}}
\end{figure}

\subsection{Source detection and X-ray photometry}
\label{sec:catalog_photometry}

We applied \textit{wavdetect} to the merged raw images in the full, soft, and hard bands using a $\sqrt{2}$ sequence of wavelet scales (1, 1.414, 2, 2.828, 4, 5.656, 8, 11.314, and 16 pixels) and a false-positive probability threshold of $10^{-6}$. We merged the three source lists using a matching radius of $3\arcsec$ and adopted positions from the full-, soft-, or hard-band detections in order of priority. This procedure produced 262 candidates.

We used \textit{srcflux} to extract X-ray photometry in the three bands. We set \texttt{regions=simple} to generate circular source regions enclosing 90\% of the PSF at 1~keV. Background annuli had inner and outer radii of 1.7 and 5 times the source-region radius. In crowded regions, \textit{roi} excluded overlaps between source regions and removed neighboring sources from the corresponding background regions. We used \texttt{psfmethod=arfcorr} to correct for the PSF fraction of each extraction region.

For each candidate, we calculated the binomial no-source probability,
\begin{equation} \label{eq:Pb}
P_{\mathrm{B}}=\sum_{X=S}^{N}\frac{N!}{X!(N-X)!}p^X(1-p)^{N-X},
\end{equation}
where $S$ and $B$ are the counts in the source and background regions, $N=S+B$, and $p=1/(1+\mathrm{BACKSCAL})$. We retained sources with $P_{\rm B}<0.01$ in at least one band, giving a final catalog of 251 sources.

\begin{deluxetable}{cc}
\tablecaption{Detection-band combinations for the X-ray sources\label{tab:detection}}
\tablecolumns{2}
\tablewidth{0pt}
\tablehead{\colhead{Band} & \colhead{Number of Sources}}
\startdata
$F + S + H$ & 199 \\
$F + S$ & 7 \\
$F + H$ & 40 \\
$F$ & 1 \\
$S$ & 3 \\
$H$ & 1 \\
\hline
Total & 251 \\
\enddata
\tablecomments{F, S, and H denote the full (0.5--7~keV), soft (0.5--2~keV), and hard (2--7~keV) bands. Each row is an exclusive detection-band combination.}
\end{deluxetable}

For detected sources, we calculated net-count uncertainties following \citet{Gehrels1986ApJ...303..336G}. For a source not detected in a given band, we assigned a 90\% confidence-level upper limit following \citet{Kraft1991ApJ...374..344K}. We estimated hardness ratios with \texttt{FastHR} \citep{Zou2023ApJ...950..136Z}. Count rates and upper limits were converted to fluxes with \textit{modelflux} and the merged response files generated by \textit{srcflux}. We adopted an absorbed power law with $N_{\rm H}=2.7\times10^{20}~{\rm cm}^{-2}$ \citep{HI4PI2016A&A...594A.116H} and $\Gamma=1.4$.

\begin{figure*}[htbp]
\centering
\includegraphics[width=3.3in]{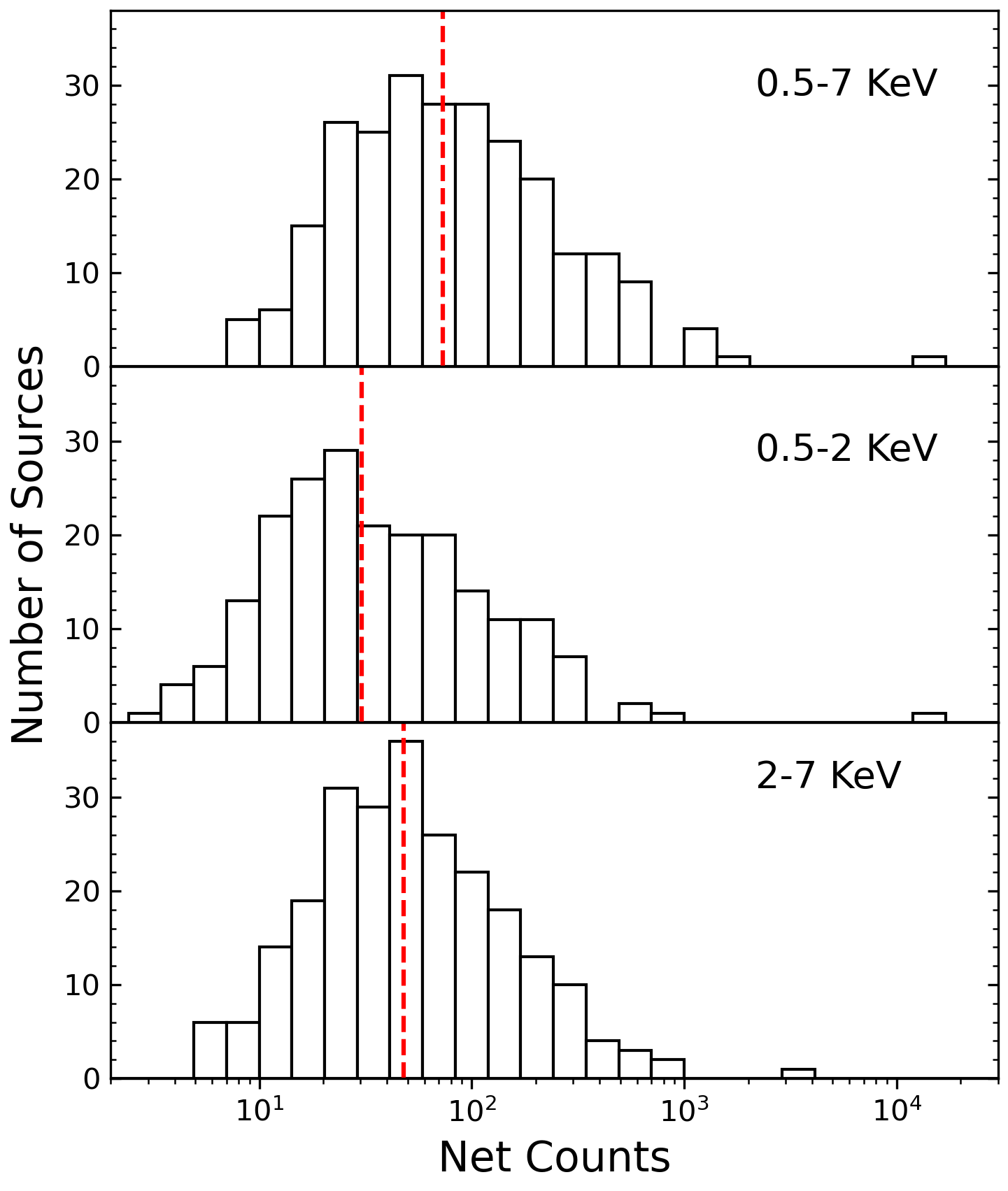}
\includegraphics[width=3.3in]{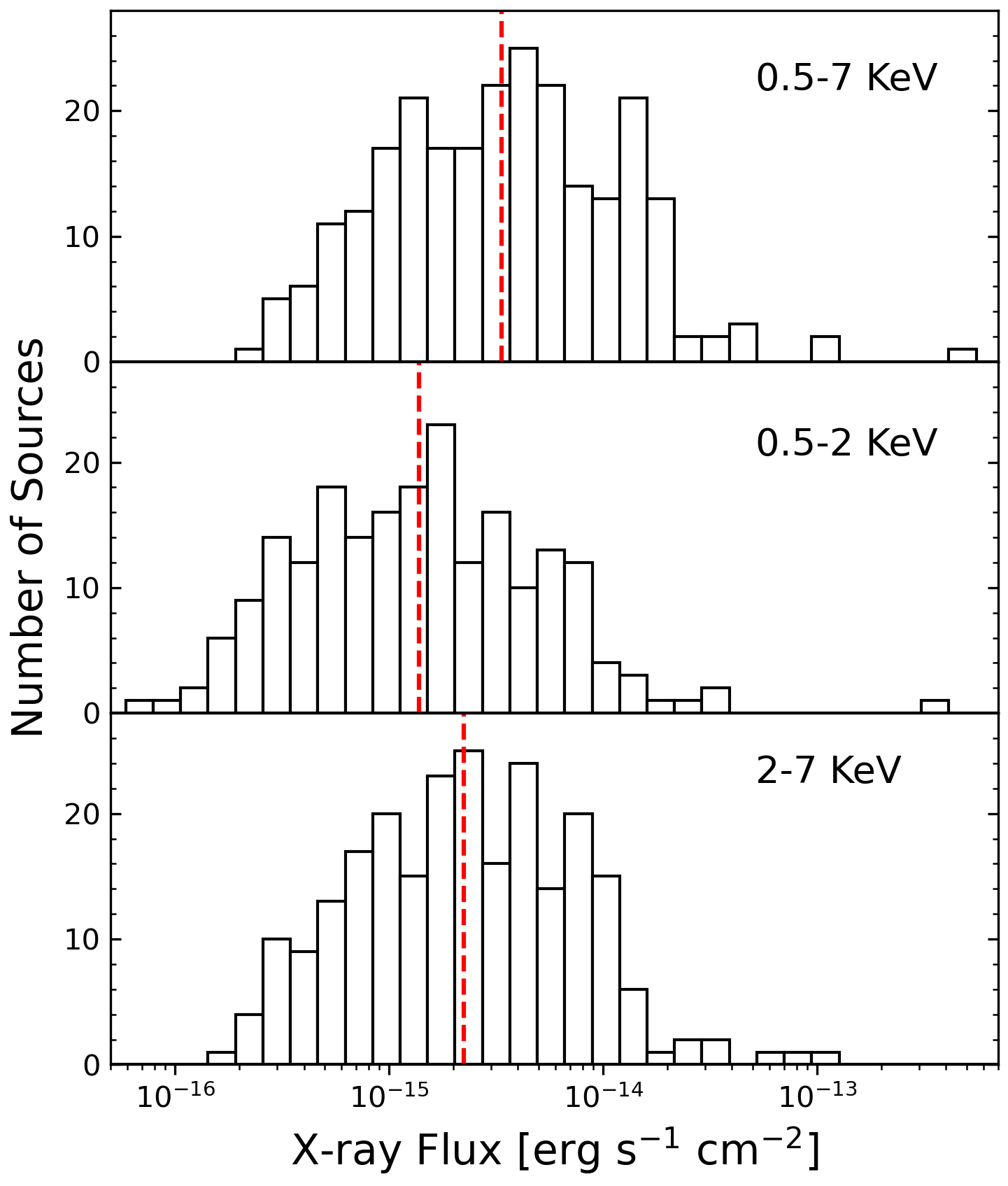}
\caption{{Left:} Net-count distributions for sources detected in the full, soft, and hard bands. The red dotted lines mark median net counts of 73.1, 30.3, and 48.0. {Right:} Aperture-corrected flux distributions for the detected sources. The red dotted lines mark median fluxes of $(3.35,1.38,2.24)\times10^{-15}~\mathrm{erg~s^{-1}~cm^{-2}}$. Sources with upper-limit estimates are excluded. \label{fig:distri}}
\end{figure*}

\subsection{Multiwavelength counterparts and catalog columns}
\label{sec:catalog_description}

We used \texttt{NWAY} \citep{Salvato2018MNRAS.473.4937S} to search for counterparts in the following catalogs:
\begin{enumerate}[noitemsep,topsep=1pt]
\item The \textit{GALEX} GR6/7 catalog, with FUV and NUV photometry \citep{Martin2005ApJ...619L...1M,Bianchi2017ApJS..230...24B}. The effective $5\sigma$ depths measured from local sources with $4.5<\mathrm{S/N}<5.5$ are approximately 20.64 and 21.10~mag.
\item The DESI Legacy Imaging Survey DR10 catalog, with $g$, $r$, $i$, and $z$ photometry and unWISE forced photometry in W1--W4 \citep{DESI2019AJ....157..168D}. The median $5\sigma$ point-source depths are $(g,r,i,z)=(26.00,24.81,23.56,23.37)$ and $(\mathrm{W1},\mathrm{W2},\mathrm{W3},\mathrm{W4})=(21.60,20.70,16.92,14.55)$.
\item The Subaru/HSC catalog, with $i2$, $z$, and NB0926 photometry. The corresponding $5\sigma$ depths are 27.31, 26.27, and 25.71~mag in $1\farcs5$-diameter apertures \citep{Wang2024ApJ...962L..11W}.
\item The VHS DR7 catalog, with $J$- and $K_s$-band photometry \citep{McMahon2013Msngr.154...35M}. The local effective $5\sigma$ depths are approximately 20.43 and 19.58~mag.
\item The JWST/NIRCam F115W, F200W, and F356W catalogs from ASPIRE, with $5\sigma$ depths of 27.51, 28.26, and 28.41~mag in $0\farcs32$-diameter apertures \citep{Wang2026ApJ}.
\item The HST/WFC3 F125W and F160W catalogs, with $5\sigma$ depths of 28.26 and 27.98~mag in $0\farcs4$-diameter apertures \citep{Protusova2025AA}.
\end{enumerate}

The primary criteria were \texttt{match\_flag}=1, a separation smaller than $3\arcsec$, and $p_{\mathrm{any}}>0.5$. These cuts follow the calibration of \citet{kim2023ApJS..268...17K}, who found false-match rates below 5\% for comparable criteria; we use this as an external benchmark rather than a field-specific measurement.

The primary matching identified at least one reliable counterpart for 227 X-ray sources (90.4\%), assigned \texttt{matched\_flag}=0. For the remaining 24 sources, we retained the most probable association within $3\arcsec$ when $p_{\mathrm{any}}>0$. This secondary step recovered plausible counterparts for 20 sources, assigned \texttt{matched\_flag}=1. XIDs~9, 32, 197, and 250 have no visible counterparts in the HSC images; we assigned HSC flux upper limits and \texttt{matched\_flag}=2.

\begin{deluxetable*}{ccccccccccccccc}
\tablewidth{0pt}
\tabletypesize{\scriptsize}
\setlength{\tabcolsep}{2pt}
\tablecaption{Main X-ray source catalog\label{tbl:maincatalog}}
\digitalasset
\tablehead{
\colhead{XID} & \colhead{R.A.} & \colhead{Decl.} & \colhead{pos\_err} & \colhead{OffAxis} & \colhead{Psf\_Exp} &
\colhead{F} & \colhead{F\_lo} & \colhead{F\_up} & \colhead{S} & \colhead{S\_lo} & \colhead{S\_up} & \colhead{H} & \colhead{H\_lo} & \colhead{H\_up} \\[-5pt]
\colhead{} & \colhead{(deg)} & \colhead{(deg)} & \colhead{(arcsec)} & \colhead{(arcmin)} & \colhead{(ks)} & \multicolumn{9}{c}{(counts)}
}
\startdata
1 & 137.74515 & $-4.34490$ & 0.33 & 6.43 & 752.0 & 240 & 224 & 255 & 99 & 88 & 109 & 141 & 128 & 153 \\
2 & 137.69275 & $-4.33495$ & 0.14 & 6.19 & 751.9 & 1120 & 1090 & 1160 & 537 & 514 & 561 & 586 & 561 & 610 \\
3 & 137.76654 & $-4.32926$ & 0.11 & 5.84 & 752.0 & 1490 & 1450 & 1530 & 814 & 785 & 844 & 673 & 646 & 699 \\
4 & 137.66023 & $-4.31172$ & 0.19 & 6.06 & 752.6 & 596 & 572 & 621 & 267 & 251 & 284 & 329 & 311 & 348 \\
5 & 137.70590 & $-4.30977$ & 0.22 & 4.50 & 749.4 & 190 & 176 & 205 & 55 & 47 & 62 & 136 & 123 & 148 \\
\enddata
\tablecomments{The table is published in its entirety in machine-readable format. A portion containing the first 15 columns and five sources is shown here. The full table contains 251 sources and 92 columns.}
\end{deluxetable*}
\onecolumngrid

Columns 1--3 give the source ID and X-ray position. Column 4 gives the formal positional uncertainty, calculated as $\sigma=R_{90}/\sqrt{C}$ \citep{Puccetti2009ApJS..185..586P}, where $C$ is the full-band net count and $R_{90}$ is the 90\% encircled-energy radius from \citet{Evans2024ApJS..274...22E}; this estimate does not include an additional systematic floor. Columns 5 and 6 give the off-axis angle and exposure weighted by the PSF. Columns 7--33 give net counts, aperture-corrected count rates, and aperture-corrected fluxes in the full, soft, and hard bands, including the corresponding bounds. For a source not detected in a band, these quantities are reported as 90\% upper limits and the bound columns are left blank. Columns 34--36 give $P_{\rm B}$ in the three bands, and Columns 37--39 give the hardness ratio and its bounds. Columns 40--91 contain the counterpart positions and photometry from JWST, HST, HSC, DESI Legacy, VHS, and \textit{GALEX}; no additional correction for Galactic foreground extinction was applied. Column 92 gives the counterpart reliability flag.

\subsection{Cumulative number counts}
\label{app:number_counts}

We used the sky-coverage curves shown in Figure~\ref{fig:sensimap} to calculate the cumulative number counts,
\begin{equation} \label{eq:NS}
N(>S)=\sum_{i=1}^{N_S}\frac{1}{\Omega_i}\ \mathrm{deg}^{-2},
\end{equation}
where $N_S$ is the number of sources brighter than $S$ with nonzero sky coverage and $\Omega_i$ is the sky coverage at the flux of the $i$th source.

\begin{figure*}[htbp]
\centering
\includegraphics[width=0.7\textwidth]{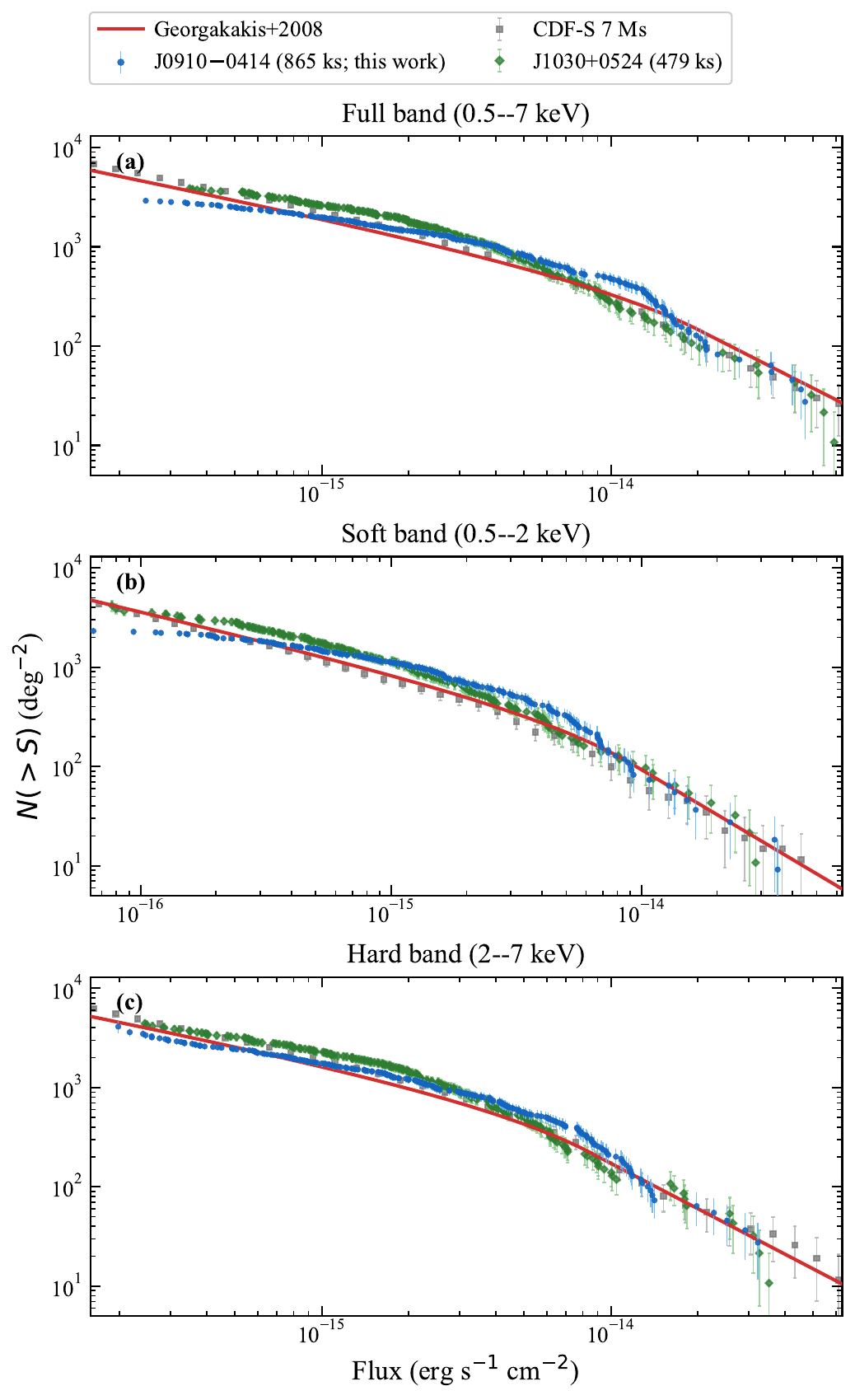}
\caption{Cumulative number counts $N(>S)$ in the full ({top}), soft ({middle}), and hard ({bottom}) bands. Blue circles show the J0910$-$0414 sources satisfying the catalog criterion and having nonzero sky coverage. Grey squares show the 7~Ms CDF-S measurements \citep{Luo2017ApJS..228....2L}, green rhombuses show the J1030+0524 measurements \citep{Nanni2020A&A}, and red curves show the broken-power-law model of \citet{Georgakakis2008MNRAS.388.1205G}, converted to the 0.5--7 and 2--7~keV bands assuming $\Gamma=1.4$ where applicable. \label{fig:NSplot}}
\end{figure*}

Figure~\ref{fig:NSplot} compares the J0910$-$0414 number counts with the empirical relation of \citet{Georgakakis2008MNRAS.388.1205G} and measurements from the 7~Ms CDF-S \citep{Luo2017ApJS..228....2L} and J1030+0524 \citep{Nanni2020A&A}. The J0910$-$0414 field lies at Galactic coordinates $(l,b)=(234.64^\circ,+28.28^\circ)$, so foreground stars may contribute to the observed point-source counts, particularly in the soft band. We do not subtract a separate stellar component.

The \citeauthor{Georgakakis2008MNRAS.388.1205G} model was derived from approximately 6300 sources across \chandra\ surveys spanning $11.8~\mathrm{deg^2}$. For the full and hard bands, we converted its original 0.5--10 and 2--10~keV relations to 0.5--7 and 2--7~keV assuming $\Gamma=1.4$, matching both the model's count-rate-to-flux conversion and the spectral assumption adopted for our catalog.

The J0910$-$0414 number counts are broadly consistent with the comparison surveys and the \citeauthor{Georgakakis2008MNRAS.388.1205G} model in all three bands. We note a modest excess around $S\sim10^{-14}~\mathrm{erg~s^{-1}~cm^{-2}}$ and lower counts at the faintest fluxes. Given the survey area, these differences may arise from field-to-field variance, foreground stars, and differences in completeness near the flux limits. We therefore do not interpret them as evidence for a different underlying X-ray source population.

\section{X-ray spectral fits and optical classifications of four hard \chandra\ sources}
\label{app:specfit}

Of the 251 X-ray sources in our catalog, 23 have spectroscopic redshifts from DESI \citep{DESI2026AJ}. For each source, we extracted the source and local-background spectra separately, together with the associated ARF and RMF response files, using the CIAO \textit{srcflux} pipeline (Section~\ref{sec:catalog_photometry}). We fitted the 0.5--7~keV spectra with \textsc{sherpa} \citep{Siemiginowska2024arXiv}, grouping them to a minimum of one count per bin and using the $W$-statistic, which treats the Poisson source and background data without subtracting the background spectrum. The Galactic neutral-hydrogen column density was fixed at $N_{\rm H,Gal}=2.7\times10^{20}~\rm cm^{-2}$ \citep{HI4PI2016A&A...594A.116H}.

As a first pass, we fitted all 23 DESI sources with \textsc{phabs}$\times$\textsc{zpowerlw}, leaving the photon index $\Gamma$ and normalization free. Most have photon indices typical of AGN \citep[$\Gamma=1.7$--2.3;][]{Scott2011MNRAS}, whereas XIDs~12, 87, 90, and 202 have apparent slopes $\Gamma_{\rm app}<1$ (Table~\ref{tab:specfit_hard}). Such hard effective spectra are a standard X-ray AGN-selection signature \citep[e.g.,][]{Luo2017ApJS..228....2L}.

To assess whether these hard spectra indicate absorption, we compared the unobscured model $U=\textsc{phabs}\times\textsc{zpowerlw}$ with the obscured model $O=\textsc{phabs}\times\textsc{zphabs}\times\textsc{zpowerlw}$. Here, \textsc{phabs} represents Galactic absorption, \textsc{zphabs} represents neutral absorption at the source redshift, and \textsc{zpowerlw} is the intrinsic AGN power-law continuum; both models fix $\Gamma=1.8$. For XIDs~12 and 87, the absorbed power-law model left clear low-energy excesses, so we added an unabsorbed copy of the same power law, $O=\textsc{phabs}\times(\textsc{zphabs}\times\textsc{zpowerlw}+f_{\rm scatt}\times\textsc{zpowerlw})$. We use this as a phenomenological scattered-power-law component: it represents a small fraction of the intrinsic nuclear continuum that reaches the observer without the line-of-sight absorption applied to the primary component and is often interpreted as Thomson scattering by Compton-thin circumnuclear gas in obscured AGN \citep{Turner1997ApJ...488..164T,Ueda2007ApJ...664L..79U,Noguchi2010ApJ...711..144N,Gupta2021MNRAS.504..428G}. Its slope is tied to that of the primary continuum and $f_{\rm scatt}$ is constrained to $0$--$0.2$; the best-fitting values are 0.13 and 0.03 for XIDs~12 and 87, respectively. The soft component is included only in $O$. XID~12 shows an additional excess in the Fe-K band. To prevent this feature from influencing the absorption test, we excluded the rest-frame 5.8--7.2~keV interval (4.41--5.47~keV observed) from both its $U$ and $O$ continuum fits. No Gaussian component was included in this comparison; the Fe feature was fitted separately after selecting the preferred continuum model.

\begin{figure*}[htbp]
\centering
\includegraphics[width=\textwidth]{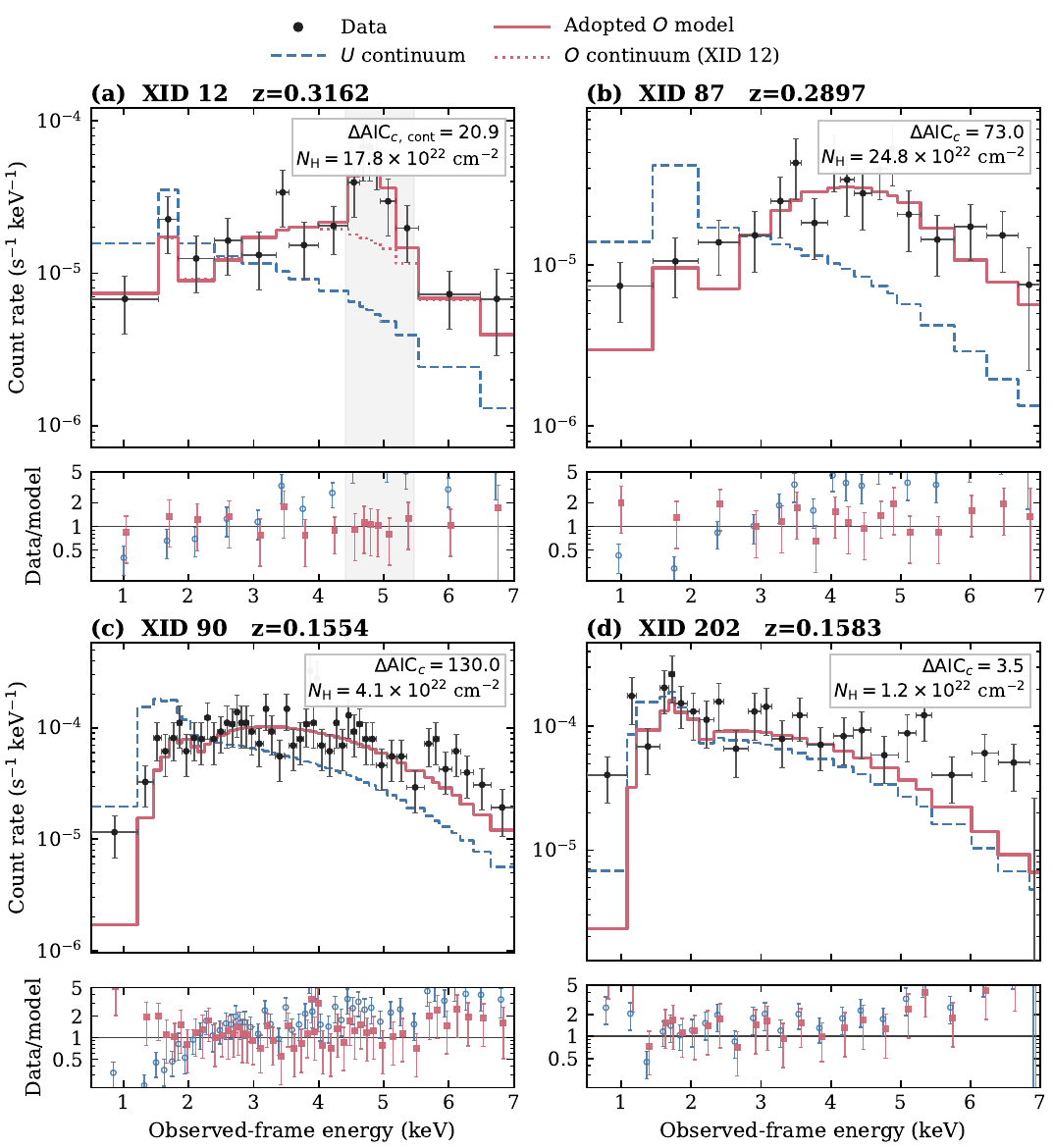}
\caption{X-ray spectra and continuum-model comparisons for the four hard DESI sources. Black points show the data with $1\sigma$ uncertainties, blue dashed curves show the canonical unobscured models ($U$), and red solid curves show the final adopted obscured models ($O$). For XID~12, the gray band marks the rest-frame 5.8--7.2~keV interval excluded when comparing the $U$ and $O$ continua; the red dotted curve is the resulting absorbed-plus-scattered continuum extrapolated across the full band, whereas the red solid curve is the full-band $O$ model after adding the free-width Fe-line Gaussian. For XID~87, the red solid curve is an absorbed plus scattered continuum; for XIDs~90 and 202, it is an absorbed power law. The lower subpanels compare the data with the blue dashed and red solid models. The XID~12 annotation gives $\Delta\mathrm{AIC_c,cont}=\mathrm{AIC_c}(U)-\mathrm{AIC_c}(O)$ from the line-masked continuum fits; the other annotations use the full fitted band. All continua adopt $\Gamma=1.8$. The spectra are grouped to a minimum of six counts per bin for display only, but are fitted with a minimum of one count per bin using the $W$-statistic and without background subtraction. \label{fig:specfit_hard}}
\end{figure*}

For each fit we calculated the small-sample Akaike information criterion,
\begin{equation}
    \mathrm{AIC_c}=W+2k+\frac{2k(k+1)}{n-k-1},
\end{equation}
where $k$ is the number of free parameters and $n$ is the number of fitted spectral bins. We define $\Delta\mathrm{AIC_c}=\mathrm{AIC_c}(U)-\mathrm{AIC_c}(O)$, so positive values favor the obscured model. The continuum comparison yields $\Delta\mathrm{AIC_c}=20.9$, 73.0, 130.0, and 3.5 for XIDs~12, 87, 90, and 202, respectively (Table~\ref{tab:specfit_hard} and Figure~\ref{fig:specfit_hard}); the XID~12 value is calculated with the Fe-K interval excluded. Thus, absorption is strongly favored for XIDs~12, 87, and 90, but only modestly favored for XID~202. This AICc comparison ranks the two specified continuum models; it is not an absolute goodness-of-fit test.

After establishing $O$ as the preferred continuum for XID~12, we restored the full 0.5--7~keV band and added an XSPEC \textsc{gaussian}, giving $O=\textsc{phabs}\times(\textsc{zphabs}\times\textsc{zpowerlw}+f_{\rm scatt}\times\textsc{zpowerlw}+\textsc{gaussian})$. The rest-frame centroid, intrinsic width, and normalization were free, with $5.8\leq E_{\rm Fe}\leq7.2$~keV, $0\leq\sigma_{\rm rest}\leq2$~keV, and non-negative normalization. The full-band obscured-plus-soft continuum without the Gaussian gives $W/{\rm dof}=66.8/82$; adding the line improves the fit to $53.9/79$ and removes the residual near 4.8~keV. The fitted centroid is $E_{\rm Fe}=6.29^{+0.11}_{-0.10}$~keV, and the best-fitting width is $\sigma_{\rm rest}=0.24^{+0.12}_{-0.15}$~keV. However, the source aperture contains only 95 counts over 0.5--7~keV, including 38 in the Fe-search interval, so the line width and equivalent width are poorly constrained. Because the centroid and width were searched over, and because the spectrum has low counts, the improvement in $W$ has not been calibrated with posterior-predictive simulations that include the look-elsewhere effect; we therefore treat this as a candidate Fe-K feature.

The absorption-corrected rest-frame 2--10~keV luminosities of the four sources span $\log(L_{\rm X}/{\rm erg~s^{-1}})=41.88$--42.34. The upper end is approximately the 2--10~keV equivalent, for $\Gamma=1.8$, of the intrinsic $L_{0.5-7\,{\rm keV}}\geq3\times10^{42}~{\rm erg~s^{-1}}$ criterion used by \citet{Luo2017ApJS..228....2L} to identify luminous X-ray AGN; the luminosities therefore do not by themselves establish the nature of the sources. The spectral evidence is substantially stronger for XIDs~12, 87, and 90: all three exhibit extremely hard apparent continua, and their spectra strongly favor absorbed over unobscured power-law models, with $\Delta\mathrm{AIC_c}=20.9$--130.0 and inferred column densities of $N_{\rm H}=(4.1$--$24.8)\times10^{22}~{\rm cm^{-2}}$. XID~12 additionally shows a candidate Fe-K feature.

We uniquely matched all four sources to the DESI DR1 AGN Host Galaxies Physical Properties value-added catalog \citep{Siudek2024A&A} and adopted its cataloged stellar masses and SFRs. The fitted stellar masses span $\log(M_\star/M_\odot)=9.56$--10.69. The nominal SFRs of XIDs~12, 90, and 202 are 3.6, 1.9, and $1.2~M_\odot~{\rm yr}^{-1}$, respectively, whereas the SFR of XID~87 is unconstrained. The stellar-mass PDF quality criterion recommended for the catalog is satisfied by all four sources, but the SFR estimate for XID~87 fails the corresponding criterion and that of XID~202 has a 1.79~dex uncertainty. Applying the redshift-dependent LMXB+HMXB relation of \citet{Lehmer2016ApJ} to the central values gives $\log(L_{2-10}^{\rm XRB}/{\rm erg~s^{-1}})=40.39$, 40.21, 39.75, and 40.26 for XIDs~12, 87, 90, and 202, respectively. These central predictions are factors of 48, 135, 238, and 42 below the corresponding absorption-corrected luminosities; the XID~87 value is effectively the LMXB contribution because its HMXB term cannot be constrained. These estimates also do not include the expected 0.2--0.4~dex galaxy-to-galaxy scatter in the relation. Conversely, reproducing the observed luminosities entirely with this XRB relation would require SFRs of approximately 420, 830, 580, and $320~M_\odot~{\rm yr}^{-1}$, respectively. The DESI spectra discussed below do not show evidence that these galaxies are undergoing star formation at this level: XIDs~87 and 202 are old, continuum-dominated systems, XID~12 is securely BPT-classified as an AGN, and only XID~90 has nominally star-forming line ratios. Although the broadband SED and optical spectra cannot exclude heavily dust-obscured star formation in the sources with poorly constrained SFRs, this large mismatch further disfavors a galaxy-wide X-ray-binary explanation. The combined X-ray and optical evidence supports obscured accretion as the most likely origin of the emission from XIDs~12, 87, and 90, which we classify as likely obscured AGN. In contrast, XID~202 has the lowest luminosity in the group and shows only modest evidence for absorption ($\Delta\mathrm{AIC_c}=3.5$); we therefore retain it more conservatively as a hard-spectrum AGN candidate without a secure obscuration classification.

We also examined the DESI spectra using the emission-line and continuum measurements from the DESI DR1 FastSpecFit value-added catalog (version 3.0; \citealt{Moustakas2023ascl.soft08005M}). We required signal-to-noise ratio ${\rm S/N}\geq3$ in H$\beta$, [O~III]~$\lambda5007$, H$\alpha$, and [N~II]~$\lambda6584$ before assigning a [N~II] BPT classification \citep{Kauffmann2003MNRAS,Kewley2001ApJ...556..121K}. XID~12 lies securely in the BPT-AGN region. XIDs 87 and 202 lack the four significant lines needed for BPT classification and have strong 4000~\AA\ breaks ($D_n4000=1.84$ and 1.83), indicating old, continuum-dominated hosts. XID~90 has nominally star-forming line ratios, but [N~II] is marginal (${\rm S/N}=2.9$), so no secure BPT class is assigned.

\begin{figure*}[htbp]
    \centering
    \includegraphics[width=\linewidth]{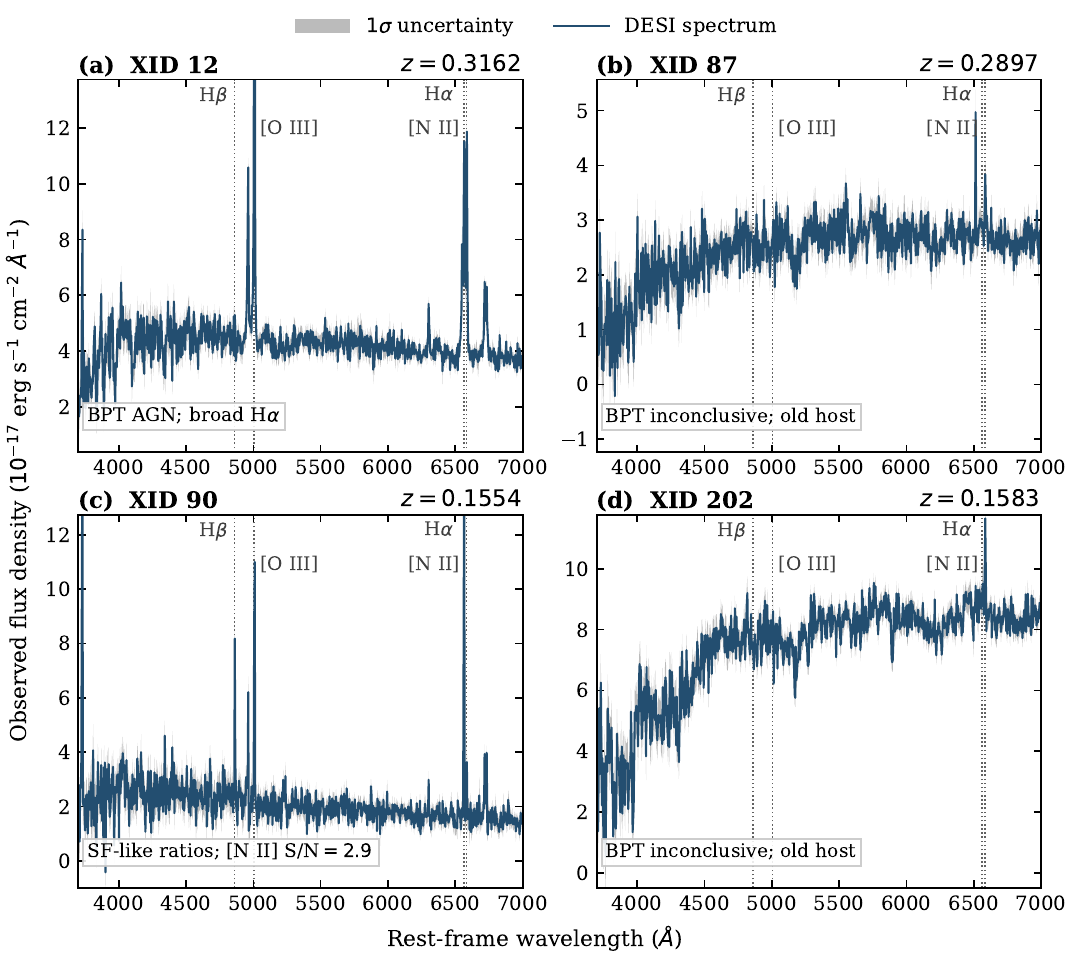}
    \caption{Rest-frame DESI optical spectra of the four hard \chandra\ sources, binned to 4~\AA\ for display. The dark-blue curve is the observed flux density and the gray band shows its $1\sigma$ uncertainty. Dotted lines mark H$\beta$, [O~III]~$\lambda5007$, H$\alpha$, and the [N~II] doublet. Panel annotations summarize the DESI DR1 FastSpecFit v3.0 diagnostics; ``BPT inconclusive'' means that one or more of the four required lines has ${\rm S/N}<3$. 
\label{fig:desispec}}
    
\end{figure*}

Consequently, these optical spectra are not a uniform set of textbook optically obscured AGN: one shows clear AGN lines, two are weak-line old galaxies, and one is host/star-formation dominated. Such differences do not contradict the X-ray inference, because optical and X-ray classifications probe different sightlines and the optical spectrum can be diluted by host-galaxy light; substantial disagreement between the two classification schemes is observed in survey AGN samples \citep{Merloni2014MNRAS.437.3550M}. Figure~\ref{fig:desispec} displays the spectra and makes the applicable diagnostic limitation explicit for each source.

\begin{deluxetable*}{lcccccccccc}
\tablecaption{Continuum-model comparison for four hard X-ray sources with DESI redshifts\label{tab:specfit_hard}}
\tablewidth{0pt}
\tabletypesize{\scriptsize}
\tablehead{
\colhead{XID} &
\colhead{$z$} &
\colhead{$\Gamma_{\rm app}$} &
\colhead{$N_{\rm H}$} &
\colhead{$W$/dof ($U$)} &
\colhead{$\mathrm{AIC_c}(U)$} &
\colhead{$O$ continuum} &
\colhead{$W$/dof ($O$)} &
\colhead{$\mathrm{AIC_c}(O)$} &
\colhead{$\Delta\mathrm{AIC_c}$} &
\colhead{$\log L_{\rm X}$} \\[-5pt]
\colhead{} & \colhead{} & \colhead{} & \colhead{($10^{22}$)} &
\colhead{} & \colhead{} & \colhead{} & \colhead{} &
\colhead{} & \colhead{} & \colhead{$(\mathrm{erg\,s^{-1}})$} \\[-5pt]
\colhead{(1)} & \colhead{(2)} & \colhead{(3)} & \colhead{(4)} &
\colhead{(5)} & \colhead{(6)} & \colhead{(7)} & \colhead{(8)} &
\colhead{(9)} & \colhead{(10)} & \colhead{(11)}
}
\startdata
12  & 0.3162 & $-0.16^{+0.25}_{-0.25}$ & $17.8^{+13.8}_{-8.9}$  & 66.6/53  & 68.7  & A+S    & 41.4/51  & 47.8  & 20.9  & $42.07^{+0.22}_{-0.19}$ \\
87  & 0.2897 & $-0.58^{+0.32}_{-0.33}$ & $24.8^{+8.4}_{-6.3}$ & 137.6/89 & 139.7 & A+S    & 60.4/87  & 66.7  & 73.0  & $42.34^{+0.12}_{-0.11}$ \\
90  & 0.1554 & $0.24^{+0.15}_{-0.15}$  & $4.1^{+0.6}_{-0.5}$   & 277.5/221& 279.5 & A      & 145.5/220& 149.5 & 130.0 & $42.13^{+0.04}_{-0.04}$ \\
202 & 0.1583 & $0.96^{+0.27}_{-0.28}$  & $1.2^{+0.7}_{-0.6}$   & 117.2/118& 119.2 & A      & 111.6/117& 115.7 & 3.5   & $41.88^{+0.08}_{-0.08}$ \\
\enddata
\tablecomments{Columns: (1) Catalog XID; (2) DESI spectroscopic redshift; (3) apparent photon index from the initial unabsorbed free-$\Gamma$ fit; (4) intrinsic column density of the obscured continuum model, in units of $10^{22}~\mathrm{cm^{-2}}$; (5)--(6) fit statistic and $\mathrm{AIC_c}$ for the canonical unobscured continuum $U$; (7) form of the compared obscured continuum $O$, where A denotes an absorbed power law and S an unabsorbed scattered power law; (8)--(9) fit statistic and $\mathrm{AIC_c}$ for continuum model $O$; (10) $\mathrm{AIC_c}(U)-\mathrm{AIC_c}(O)$; (11) absorption-corrected rest-frame 2--10~keV luminosity inferred from continuum model $O$, in $\log(\mathrm{erg\,s^{-1}})$. Quoted uncertainties on $\Gamma_{\rm app}$, $N_{\rm H}$, and $\log L_{\rm X}$ are 68\% confidence intervals. Both $U$ and $O$ fix the intrinsic continuum at $\Gamma=1.8$, and the soft component is included only in $O$ for XIDs 12 and 87. For XID~12, columns (4)--(10) are derived from continuum fits that exclude the rest-frame 5.8--7.2~keV interval; the Fe line is fitted separately after the continuum comparison and is therefore not included in the tabulated $O$ statistics.}
\end{deluxetable*}
\onecolumngrid

\bibliography{sample701}{}
\bibliographystyle{aasjournalv7}

\end{document}